\documentclass[%
 reprint,
 amsmath,amssymb,
 aps,
 pre
]{revtex4-1}

\usepackage{graphicx}
\usepackage{dcolumn}
\usepackage{bm}
\usepackage{enumerate}
\usepackage{enumitem}

\begin{document}

\title{Delta-Davidson method for interior eigenproblem in many-spin systems}

\author{Haoyu Guan}
\affiliation{%
 School of Physics and Technology, Wuhan University, Wuhan, Hubei 430072, China
}%
\author{Wenxian Zhang}%
 \email{wxzhang@whu.edu.cn}
\affiliation{%
 School of Physics and Technology, Wuhan University, Wuhan, Hubei 430072, China
}%


\date{\today}

\begin{abstract}
Many numerical methods, such as tensor network approaches including density matrix renormalization group calculations, have been developed to calculate the extreme/ground states of quantum many-body systems. However, little attention has been paid to the central states, which are exponentially close to each other in terms of system size. We propose a Delta-Davidson (DELDAV) method to efficiently find such interior (including the central) states in many-spin systems. The DELDAV method utilizes Delta filter in Chebyshev polynomial expansion combined with subspace diagonalization to overcome the nearly degenerate problem. Numerical experiments on Ising spin chain and spin glass shards show the correctness, efficiency, and robustness of the proposed method in finding the interior states as well as the ground states. The sought interior states may be employed to identify many-body localization phase, quantum chaos, and extremely long-time dynamical structure.

\end{abstract}

\maketitle


\section{Introduction}
\label{sec:intro}
Computation of eigenvalues and eigenstates of quantum spin systems is an important task in quantum information and condensed matter physics~\cite{QCQI, QPT}. With the full spectrum known, one can trivially explore many interesting properties of the physical systems, such as thermodynamic properties, quantum phase transitions, and evolution dynamics~\cite{IQSM, QPTT, PhysRevB.82.161308, Dob}. To solve the eigenproblem, direct numerical diagonalization or naively solving the time-independent Schr\"{o}dinger equation of the quantum many-spin systems is typically unfeasible, as physical systems of interest often involve a huge Hilbert space whose dimension grows exponentially with system size. Even for a modest number of spins, say 20, the full exact diagonalization requires excessive storage and computing time.

Instead of obtaining the full spectrum, one may be only interested in a set of eigenvalues or eigenstates for parts of the spectrum in a certain energy range. For example, ground state is usually enough to identify a quantum phase transition~\cite{QPT, PhysRevLett.104.137204}. Due to its great importance and wide application in condensed matter physics, many numerical methods, such as quantum Monte Carlo~\cite{PhysRevD.27.1304, PhysRevB.43.5950}, the traditional density matrix renormalization group (DMRG) method~\cite{PhysRevLett.69.2863} and matrix product states (MPS)~\cite{PhysRevLett.93.040502} have been developed to solve the ground states of quantum many-body systems, the latter two are for one-dimensional systems while higher dimensional tensor network states~\cite{PhysRevLett.99.220405, Verstraete2008Matrix} are well-established for some two-dimensional and ladder systems. It is known that the efficiency of the tensor network approaches are limited to the systems whose ground states must obey the area law~\cite{ORUS2014117, RevModPhys.82.277}. For the quantum Monte Carlo method, it suffers from the widely known sign problem, which often requires an exponential amount of computational resources to obtain a reasonable accuracy~\cite{PhysRevB.41.9301, PhysRevLett.94.170201}. In practice, the imaginary time propagation and the Lanczos method~\cite{Lanczos2018An} are also widely used to find the ground states. However, little attention has been paid to the interior eigenvalues and eigenstates.

The interior eigenstates are useful in understanding the universality of the entanglement entropy~\cite{PhysRevLett.121.220602}, the knowledge of eigenstates near the many-body mobility edge helps to locate many-body localization transition~\cite{PhysRevLett.113.107204,mblapplication}, and the level spacing statistics in the central spectrum could distinguish quantum chaos from many-body integrable phases~\cite{PhysRevLett.52.1,sgs}. It has been shown that both typical excited eigenstates for quadratic Hamiltonians and eigenstates in the thermal phase exhibit a volume law; i.e., the von Neuman entropy of the reduced density matrix of a
subsystem scales with the subsystem’s volume~\cite{PhysRevLett.121.220602, PhysRevLett.113.107204}. Thus, the tensor network approaches are not suitable for solving the interior eigenproblem.
To compute the interior eigenstates, a strategy of matrix spectroscopy is often invoked~\cite{Thomas1980The,PhysRevE.51.3643,PhysRevE.56.4837}. It aims at finding a small set of eigenstates near a certain energy $\lambda$. The essential idea may be illustrated by the shift-invert method~\cite{TSA}, which casts the eigenvalues close to $\lambda$ to the edges of the spectrum of $G$ through a spectral transformation $G=(\mathcal H-\lambda I)^{-1}$, where $I$ is the identity matrix and we drop explicitly writing it henceforth. In this way, the original problem is transformed to finding the extreme eigenstates. However, this method suffers the rapid scaling of resources~\cite{scipost} and alternative methods without matrix factorization are explored and discussed below.

Several methods, requiring only matrix-vector products,  have been designed to compute interior eigenstates. The Davidson method~\cite{Davidson1989Super} is a preconditioned version of the Lanczos method, but it can only be effective when the matrix is nearly diagonal~\cite{YSaad}.  Similarly, the harmonic Davidson method is based on a spectral transformation utilizing the harmonic Ritz values, with iterative subspaces of rather high dimension~\cite{Dorando2007Targeted,Jordan2012Fast}. There is also the filter-diagonalization method, which uses short-time propagation with Fourier energy filtering at the desired spectrum~\cite{Neuhauser1990Bound, Neuhauser1991Time,Santra2000Parallel}. However, the time propagation of states is not necessarily needed and may be replaced by the spectral filters~\cite{PhysRevE.62.4351}.
Hybrid techniques may enhance the efficiency of the methods mentioned above.
The idea of hybridizing the Lanczos method or the Davidson method with a spectral filter is thus proposed, where the filter is designed to tune out the undesirable portion of the spectrum, while amplifying the desired portion~\cite{PhysRevE.51.3643}. Following this idea, several methods have been constructed. The two layer Lanczos-Green function iteration algorithm applies the Dyson expansion of the Green's function to a Lanczos vector, while running the risk of possible divergence of the Dyson expansion~\cite{PhysRevE.51.3643}. The Chebyshev filter diagonalization method applies the Chebyshev expansion to the rectangular window function, which may become inefficient in regions with high density of states~\cite{Pieper2016High}. Recently a thick-restart Lanczos algorithm with polynomial filtering techniques has been also proposed~\cite{AFLP, TRA}.

In this paper, we propose the DELDAV method that couples the idea of Davidson method with the Dirac delta function, for simultaneous computation of the eigenvalues and eigenstates. While it belongs to the matrix spectroscopy method, the DELDAV focuses on the near-degenerate problem (high density of states). We employ the Dirac delta function filter (Delta filter) $\delta(\mathcal H-\lambda)$ to map the eigenvalues near $\lambda$ to very large positive values. Such a Delta filter has the advantage of efficiently damping the unwanted part of the spectrum and thus greatly accelerating the convergence. The employed Davidson-type methods provide remarkable flexibility in augmenting the basis with new vectors, keeping both the subspace dimension and the reorthogonalization cost small (compared with Lanczos-type methods)~\cite{cd2}. For real problems in many-spin systems, the DELDAV method is a very efficient and robust tool. Test cases are presented, with finding $10$ eigenstates at the highest density of states region for the Hilbert space dimension up to $10^6$.

The remainder of the paper is organized as follows. We briefly review the Chebyshev-Davidson method in Sec. \ref{sec:CD}, to introduce the basic idea of the filtration and the subspace diagonalization. The formalism of the DELDAV method and its applications to the quantum spin models are given in detail in Sec. \ref{sec:DD}. In Sec. \ref{sec:result}, we describe two specific many-spin models and present the results of our numerical experiments. A conclusion is given in Sec. \ref{sec:con}.

\section{Review on Chebyshev-Davidson method}
\label{sec:CD}
\begin{figure}[b]
\includegraphics[width=3.25 in]{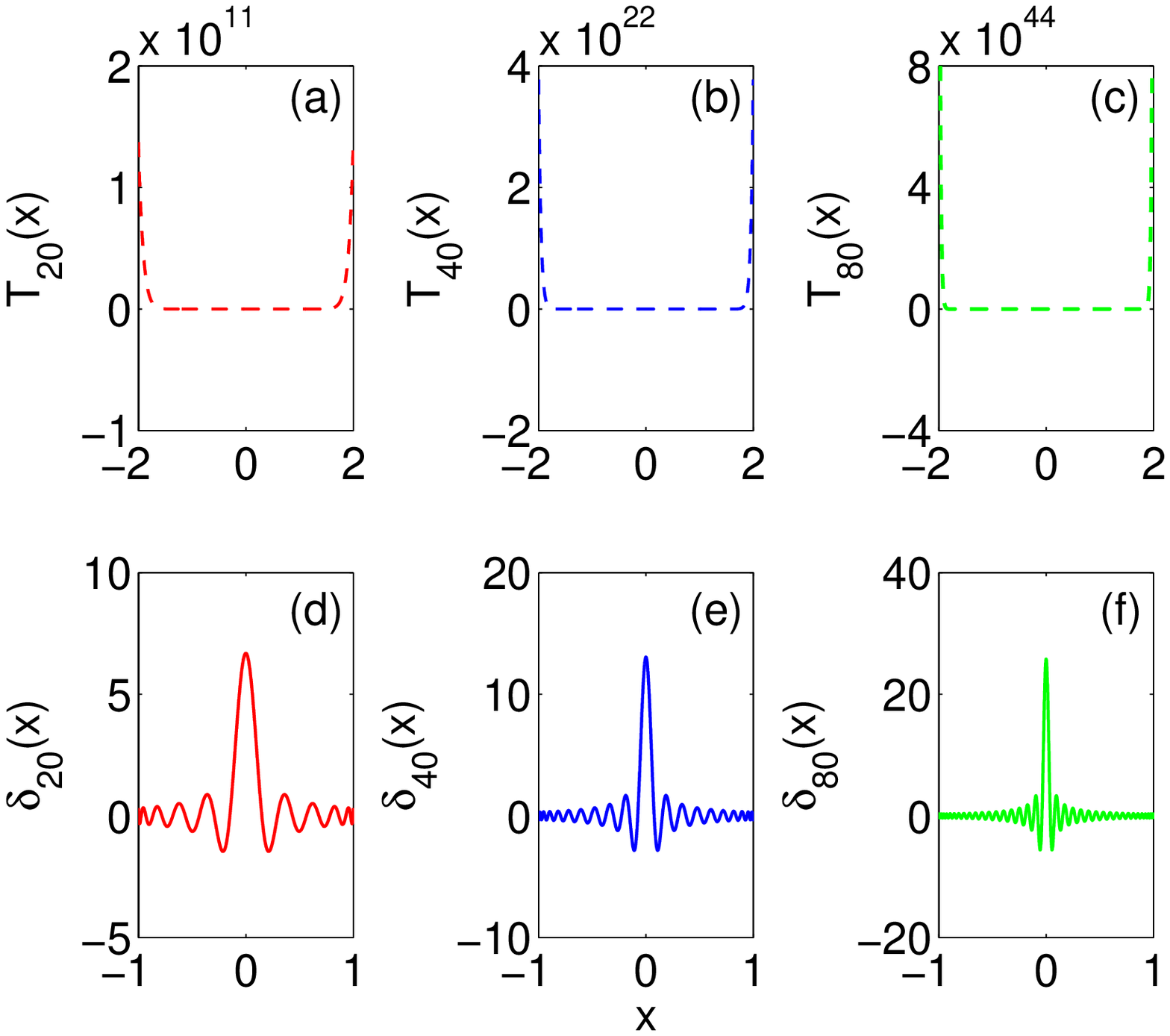}
\\
\includegraphics[width=3.25 in]{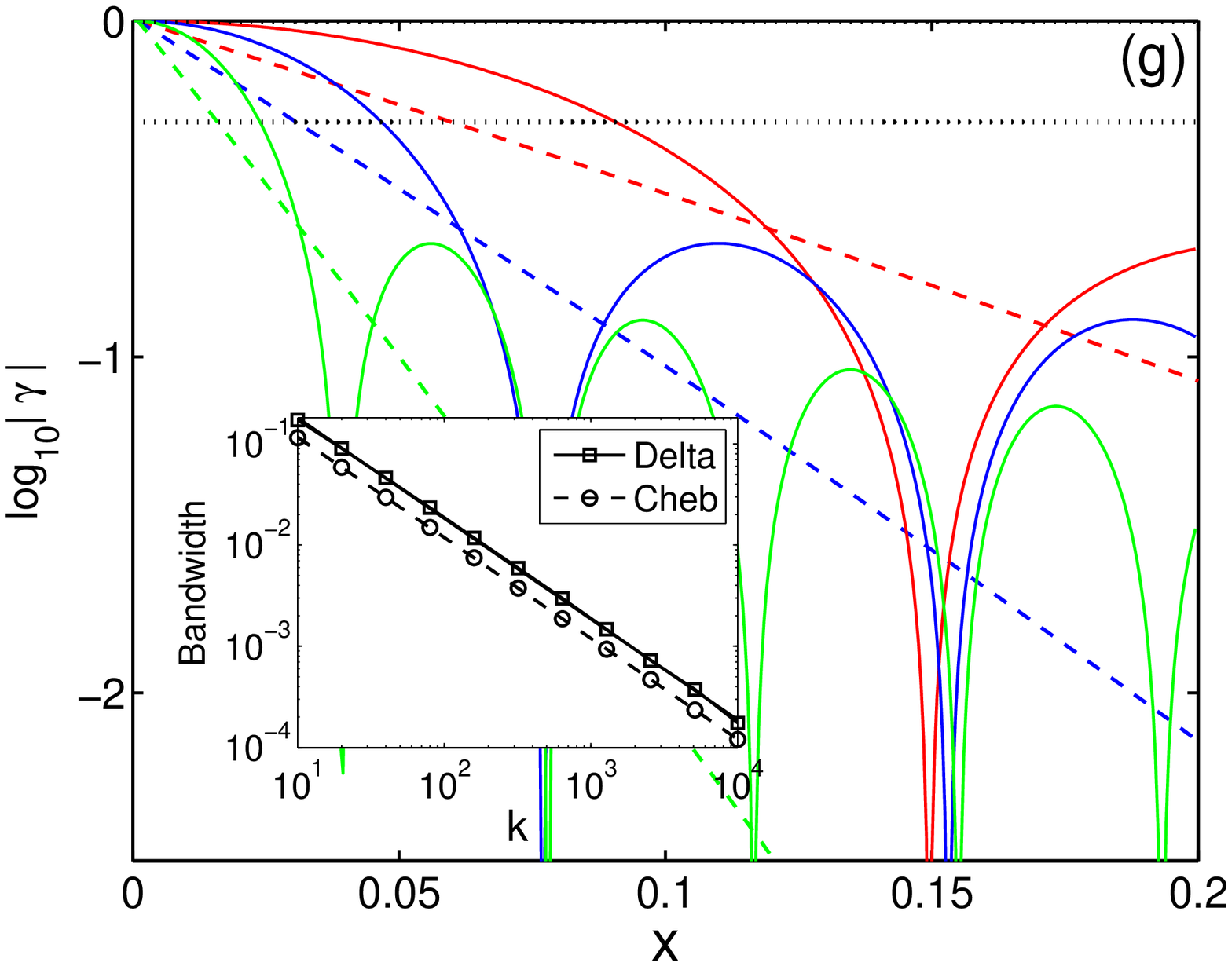}
\caption{\label{fig:shape} (Color online.) Chebyshev filter $T_k(x)$ for $k=20$ (a), $40$ (b), and $80$ (c) and Delta filter $\delta_k(x)$ (the $k$th order Chebyshev expansion of the Dirac delta function) for $k=20$ (d), $40$ (e), and $80$ (f). Note the spikes both outside $[-1, 1]$ of the Chebyshev filter and around the center of the Delta filter. (g) Comparison of the damping effect of the Chebyshev filters (dashed lines) and the Delta filters (solid lines) with polynomial orders $k=20$ (red lines), $40$ (blue lines), and $80$ (green lines). For the Chebyshev filter $\gamma=T_k(-2+x)/T_k(-2)$ and for the Delta filter $\gamma = \delta_k(x)/\delta_k(0)$. Both filters decay roughly in an exponential form. The horizontal black dotted line denotes the half maximum of the filters. Inset reveals the inverse relation between $k$ ($x$-axis) and half width at the half maximum ($y$-axis) for the Delta filters (solid line with squares) and the Chebyshev filters (dashed line with circles).}
\end{figure}

The Chebyshev-Davidson (CD) method is a Davidson-type subspace iteration using Chebyshev polynomial filters for a large symmetric/Hermitian eigenvalue problem~\cite{cd1, cd2}. This method combines the acceleration power of the Chebyshev filtering technique and the flexibility and robustness of the Davidson approach.

Typical filter in the CD method is based on Chebyshev polynomials. The $k$th order Chebyshev polynomial of the first kind is defined by
\begin{equation}
T_k\left( x \right) =\left\{ \begin{array}{l}
	\cos\left( k \cos^{-1}\left( x \right) \right) , \ \ \ \ \ \ \ \ \ \ \ \ \ |x| \le 1\ \ ,\\
	\cosh\left( k \cosh^{-1}\left( x \right) \right) ,\ \ \ \ \ \ \ \ \ \ \ \ x>1\ \ ,\\
	\left( -1 \right) ^k\cosh\left( k \cosh^{-1}\left( -x \right)\right) ,\ x<-1,\\
\end{array} \right.
\label{eq:cheb}
\end{equation}
with initial conditions $T_0(x)=1$ and $T_1(x)=x$~\cite{Handscomb2003Chebyshev}. Note that the higher order of the polynomials can be efficiently determined by using the 3-term recurrence
\begin{equation}
T_{k+1}(x)=2xT_k(x)-T_{k-1}(x).
\label{eq:rec}
\end{equation}
A remarkable property of the Chebyshev polynomial is its rapid growth outside the interval $[-1,1]$, as illustrated in Fig.~\ref{fig:shape}(a-c) and in~\cite{cd1}.

For a Hamiltonian $\mathcal H$ with a spectrum bounded in $[E_{min}, E_{max}]$, where $E_{min}$ is the minimum energy and $E_{max}$ the maximum energy, a Chebyshev filter is designed to amplify the components of the eigenstates corresponding to eigenvalues in the interval $[E_{min},a]$ and to simultaneously dampen those in the interval $[a, E_{max}]$, provided $a>E_{min}$~\cite{cd1}. To satisfy this goal, one only needs to map $[a, E_{max}]$ into $[-1, 1]$ by shift and normalization,
\begin{equation}
\label{eq:G}
\mathcal{G}=\frac{\mathcal{H}-E_c}{E_0},
\end{equation}
where $E_c=\frac{1}{2}(E_{max}+a)$ and $E_0=\frac{1}{2}(E_{max}-a)$.

The effect of Chebyshev filtering is
\begin{equation}
\label{eq:chevo}
\begin{aligned}
\left| \psi \left( k \right) \right>
&=T_k\left( \mathcal{G} \right) \left| \psi \left( 0 \right) \right> \\
&\sim\frac{1}{2}\sum_{i}{ \left( e^{k\theta _i}+e^{-k\theta _i} \right) c_i\left| \phi _i \right>},
\end{aligned}
\end{equation}
with $\left| \psi(0) \right>$ a random initial state, $\left| \phi_i \right>$ the $i$th eigenstate, $c_i$ the random coefficients of the initial state, $E_i$ the $i$th eigenvalue of $\mathcal{H}$, and 
\begin{equation}
\theta_i=\cosh^{-1}(\frac{E_c-E_i}{E_0})
\end{equation}
the order-preserving mapped eigenvalues. If $k$ is large enough, the filtration in Eq.~(\ref{eq:chevo}) effectively represents an imaginary time propagation ( i.e.
$\left| \psi \left( \tau \right) \right>
=e^{-\mathcal{H}\tau}\left| \psi \left( 0 \right) \right>
=\sum_{i}{ e^{-\tau E_i} c_i\left| \phi _i \right>}$
) at long time limit, where $k$ plays the role of imaginary time $\tau$ and $\theta_i$ the deformed eigenvalues.

After the Chebyshev filtering, one may construct the subspace as follows.
\begin{eqnarray}
\label{eq:subspace}
\mathcal{K}_d&=&\text{span} \{\left| \psi\right>, T_k(\mathcal{G}_1)\left| \psi\right>, T_k(\mathcal{G}_2) T_k(\mathcal{G}_1)\left| \psi\right>, \cdots, \nonumber \\
&& \Pi_{i=1}^{d-1}T_k(\mathcal{G}_i)\left| \psi\right> \},
\end{eqnarray}
where $\left| \psi\right>$ is usually a randomly initialized state and $\mathcal{G}_i=\mathcal{G}$, with the boundary $a$ is adjusted appropriately for each $\mathcal{G}_i$~\cite{cd1}. Whenever a new state is generated, one shall orthonormalize it against all existing states, in order to keep the $d$ states in $\mathcal{K}_d$ as an orthonormalized basis. All one needs next is to calculate the representation matrix $R$ of the Hamiltonian in this special subspace and to diagonalize it directly. The eigenstate corresponds to the largest eigenvalue of $R$ is then utilized to construct a better approximation of the ground state of the original Hamiltonian $\mathcal{H}$.

\section{Delta-Davidson method}
\label{sec:DD}

Although successful in finding the extreme states and many nearby states, the CD method is unable to find the interior states that are far from the extreme ones. 
In fact, the CD method employs either a low-pass filter or a high-pass filter. To find the interior states, one actually needs a band-pass filter. We thus introduce the Delta filter, which is obviously band-passing, to replace the Chebyshev filter.

We construct a Delta filter $\delta(\mathcal{H}-\lambda)$ to amplify the components of the eigenstates corresponding to an energy close to $\lambda$. In other words, the original interior spectrum of $\mathcal{H}$ is transformed to the extreme spectrum of $\delta(\mathcal{H}-\lambda)$. The delta function is a natural choice to do this, for it possesses a rapid growth in an infinitely small region centered at $\lambda$. Contrast to the Chebyshev filter which aims at amplifying the extreme region, the Delta filter may well amplify the extreme or interior region by adjusting $\lambda$ appropriately.

Many other filters, like Green's function filter $(\mathcal{H}-\lambda)^{-1}$~\cite{PhysRevE.51.3643,PhysRevE.56.4837} or Gaussian filter $\exp[-(\mathcal{H}-\lambda)^2]$, may do the same job. However, considering the highly near-degenerate central states in a many-spin system where the Hilbert space dimension may reach several millions, we need an extremely narrow band filter, where the band width is defined as the full width at half maximum. In fact, the narrower the band width, the better the filter effects, because a smaller band width filter dampens the unwanted part more rapidly. On the other hand, the most efficient expansion of a nonperiodic function among all polynomials is given by Chebyshev polynomial~\cite{ChebandFourier}. In consideration of this fact, we numerically find that the Delta filter works the best (having the smallest band width) among the above filters under the same order of Chebyshev expansion.

Besides, there is also a straightforward transformation $(\mathcal{H}-\lambda)^2$ being widely used to cast the interior spectrum to  extreme. In fact, this is nothing but the second order Chebyshev expansion of the delta function, ignoring the constant term. However, this transformation is not a good candidate solution when facing the highly near-degenerate problem. After the transformation, it is much harder to distinguish those neighboring eigenstates. For example, an original level spacing $10^{-7}$ of $\mathcal{H}$ is transformed to $10^{-14}$ of $\mathcal{H}^2$ (suppose $\mathcal{H}$ is rescaled to unitless for simplicity). Whereas with the Delta filter one can choose a proper expansion order to fit in a tiny region of spectrum and to significantly amplify the level spacings at the same time, as shown in Fig. \ref{fig:shape}(d-f). Below, we describe the specific details of the application of Chebyshev polynomial expansion to the Delta filter.

To efficiently expand the delta function in Chebyshev polynomials, it is necessary to limit $x\in [-1, 1]$. This region is contrast to the Chebyshev filter, which requires $x \notin [-1, 1]$ to exponentially amplify the components of the ground states. Thus, the original Hamiltonian $\mathcal{H}$ needs to be shifted by $E_c$ and be rescaled by $E_0$ for the Delta filter, where $E_c=\frac{1}{2}(E_{max}+E_{min})$ and $E_0=\frac{1}{2}(E_{max}-E_{min})$,  and note that the definitions of $E_c$ and $E_0$ are different from Eq.~(\ref{eq:G}). In this way, the rescaled operator
 \begin{equation}
 \mathcal{G}=\frac{\mathcal{H}-E_c}{E_0}
 \end{equation}
 is definitely bounded by $-1$ and $1$. For quantum spin systems, the Hamiltonian $\mathcal{H}$ is bounded both from above and from below, thus the rescaled operator $\mathcal{G}$ can readily be found.

The Chebyshev polynomial expansion of the Delta filter now becomes
\begin{equation}
\delta(\mathcal{H}-\lambda)=\frac{1}{E_0}\delta(\mathcal{G}-\lambda^{\prime})=\frac{1}{E_0}\sum_{k=0}^{\infty}{c_k(\lambda^{\prime})T_k(\mathcal{G})},
\label{eq:expan}
\end{equation}
with $\lambda^{\prime}=(\lambda-E_c)/E_0$ and we ignore the constant factor $E_0$ in Eq. (\ref{eq:expan}) henceforth.
The expansion coefficients $c_k(\lambda^{\prime})$ can be calculated using the orthogonal property of the first kind Chebyshev polynomials,
\begin{equation}
c_k\left( \lambda^{\prime} \right) =\frac{a_k}{\pi}\int_{-1}^1{\frac{T_k\left( x \right) \delta \left( x-\lambda^{\prime} \right)}{\sqrt{1-x^2}}}\text{d}x=\frac{a_kT_k\left( \lambda^{\prime} \right)}{\pi \sqrt{1-\lambda^{\prime2}}},
\label{eq:coef}
\end{equation}
where $a_k=1$ for $k=0$ and $a_k=2$ for $k\ge 1$. With the initial conditions $T_0(\mathcal{G})=1$ and $T_1(\mathcal{G})=\mathcal{G}$, the $k$th order Chebyshev polynomial can be efficiently determined using the recurrence relation Eq.~(\ref{eq:rec}). The filtered state $\left| \psi^{\prime}\right>=\delta(\mathcal{G}-\lambda^{\prime})\left| \psi\right>$ can be calculated by summing successively the terms of the series Eq.~(\ref{eq:expan}) until a predefined value $K$ of $k$ is reached. Note that as shown by different curves in  Fig.~\ref{fig:shape}(d-f), a larger $K$ corresponds to a smaller filter band width. As shown in the inset of Fig.~\ref{fig:shape}(g), the band widths of both the Chebyshev filters and the Delta filters are close to each other and inversely proportional to the expansion order $k$.

However, we remark that in practice the highly near-degenerate eigenvalues still remains a severe problem, even with  the help of Delta filtering. Generally speaking, in a quantum $N$-spin system the energy bounds grow polynomially with $N$ while the number of eigenvalues grows exponentially, thus the level spacings roughly decrease exponentially as $N$ increases~\cite{PhysRevE.62.3504}. For a typical Ising system with a size $N=20$, the level spacings of the rescaled operator $\mathcal{G}$ at the central region may be as small as $10^{-7}$ (note that $\mathcal{G}$ is unitless and bounded by $-1$ and $1$). This is not surprising, for there are $10^6$ eigenvalues inside $[-1,1]$, giving an average level spacing $2\times 10^{-6}$. To amplify the components of the wanted eigenstates and dampen the others, the cutoff expansion order $K\simeq10^7$ is required. Such a requirement already implies that $10^7$ matrix-vector operations are needed, not to mention the potential necessity of the repeated filtering iterations. Therefore, for 20-spin systems it is almost impossible to find the central region of spectrum by the Delta filtering alone.

The subspace diagonalization, combined with the Delta filtering technique, resolves this problem. Since it is hard to cope with the extremely tiny level spacing, one may amplify a cluster of states simultaneously (corresponding to a larger bandwidth) to relax the requirement of large $K$. For example, to amplify 10 states simultaneously requires $K\simeq10^6$ instead of $K\simeq10^7$ in the above case. Note that after the Delta filtering, the state becomes more concentrated in a small energy interval and is closer to the true eigenstate.
Furthermore, there is an elegant way to extract additional eigenvalue information from this filtering sequence of states, as indicated by the faster convergence of the Lanczos method compared to the power method~\cite{YSaad}. We then construct the Krylov subspace $\mathcal{K}_d$, which is formed by a set of states after iterations of the Delta filtering, as follows
\begin{eqnarray}
\mathcal{K}_d&=&\text{span} \{\left| \psi\right>, \delta_K(\mathcal{G}-\lambda^{\prime})\left| \psi\right>,\delta_K^2(\mathcal{G}-\lambda^{\prime})\left| \psi\right>, \cdots, \nonumber\\
 && \delta_K^{d-1}(\mathcal{G}-\lambda^{\prime})\left| \psi\right>\},
\end{eqnarray}
with
$
\delta_{k}(\mathcal{G}-\lambda^{\prime})= \sum_{i=0}^{k}c_i(\lambda^{\prime})T_i(\mathcal{G}) 
$
and $\delta_K\equiv\delta_{k=K}$. The base states of the Krylov subspace are linearly independent to each other and the subspace dimension $d$ is much smaller than the whole Hilbert space dimension $D$, $d\ll D$. The matrix representation $R$ of the shifted Hamiltonian $\mathcal{H}-\lambda$ in this subspace can be directly calculated and diagonalized, as further
detailed below. After the diagonalization of $R$, one can pick up a set of states that are closer to the desired answers and this is exactly the additional eigenvalue information behind the filtering sequence. Also note that for a small subspace dimension $d$ the extra efforts (to orthonormalize basis of $\mathcal{K}_d$ and diagonalize $R$) are negligible. 

To summarize, simultaneously amplifying a cluster of states reduces the order $K$ while the subspace diagonalization scales down the iteration number of filtering, leading to a tremendous speed-up of the convergence (see also the Appendix~\ref{sec:appen}). The subspace diagonalization is applicable to the degenerate, near-degenerate and non-degenerate cases. 
In this sense, the subspace diagonalization is an essential ingredient of the DELDAV method.

By combining the Delta filter and the subspace diagonalization, the explicit algorithm of the DELDAV method is as follows.
\renewcommand\theenumi{\roman{enumi}}
\renewcommand\labelenumi{(\theenumi)}
\begin{enumerate}[fullwidth,itemindent=1em]
	\item Set the cutoff order $K\sim\rho k_{want}$ for the Delta filter, where $\rho$ is the local density of states (DOS) at $\lambda$ and $k_{want}$ the number of required eigenstates. Choose the subspace dimension $d$ and the maximum iteration step $n_{max}$. We assume the whole Hilbert space dimension is $2^N$.
\item
Choose an initial random normalized trial state $\left| \psi_0\right>$, set $\left| \phi_0\right>=\left| \psi_0\right>$, $V_0=[\left| \psi_0\right>]$ and $W_0=[\left| \phi_0\right>]$.

\item
For $n=1,2,\cdots,n_{max}$, do the following iteration steps to refine the trial state.
\begin{enumerate}[fullwidth,itemindent=1em]
\item Generate a new normalized trial state through the Delta filtering
	\begin{equation}
	\left| \psi_n\right>=\beta\delta_{K}(\mathcal{G}-\lambda^{\prime})\left| \psi_{n-1}\right>,
	\end{equation}
	where $\beta$ is a normalization factor.  Then construct the $2^N\times(n+1)$ matrix $V_n$ as 
	\begin{equation}
	V_n=[V_{n-1}, \left| \psi_n\right>],
	\end{equation}	
	and note that $\text{span}(V_n)=\mathcal{K}_{n+1}$.
	\item	
	Orthonormalize $\left| \psi_n\right>$ against the base states of the orthonormal matrix $W_{n-1}$, which results in the state $\left| \phi_n\right>$. With $\left| \phi_n\right>$ we construct the matrix $W_n$ as 
	\begin{equation}
	W_n=[W_{n-1}, \left| \phi_n\right>].
	 \end{equation}	
	 The orthogonalization is performed by the method developed by Daniel-Gragg-Kaufman-Stewart, which has an appealing feature of being numerically stable~\cite{10.2307/2005398}.
	 
	 \item
	 Compute the subspace projection matrix $R_n$ of the shifted Hamiltonian
	 \begin{equation}
	 R_n=W_n^{\dag}(\mathcal{H}-\lambda)W_n.
	 \end{equation}
	 
	 Then compute the eigen-decomposition of $R_n$ as
	 \begin{equation}
	 R_nY_n=Y_n\Lambda_n,
	 \end{equation}
	 with $\Lambda_n$ a diagonal matrix and $Y_n$ storing the normalized eigenstates of $R_n$. Sort the eigenpairs of $R_n$ in non-decreasing order by the absolute value of eigenvalues.

	 \item
	 Refine the basis matrix $W_n$ by subspace rotation
	 \begin{equation}
	 W^{\prime}_n=W_nY_n,
	 \end{equation}
	 and set $W_n=W^{\prime}_n$. Since $Y_n$ is unitary, $W^{\prime}_n$ remains an orthonormal matrix. After the subspace rotation, the first state of $W_n$ is the best approximation of the eigenstate with eigenvalue nearest to $\lambda$, among all the states in the subspace $\mathcal{K}_{n+1}$.
	 
	 \item 	 
	 If the dimension of $W_n$ equals the parameter $d$, throw away the last few states of $W_n$. 
	 Then check if the first few states of $W_n$ are converged and count $m$, the number of converged states .
	 If $m\ge k_{want}$ or $n=n_{max}$, stop the iteration loop, otherwise continue it. 
\end{enumerate}

\end{enumerate}

The DELDAV method is applicable to find both the extreme and the interior states. For the extreme states, we estimate the $E_{max}$ and $E_{min}$ in the beginning, by employing the upper-bound-estimator, which costs little extra computation and bounds up the largest absolute eigenvalue~\cite{ZHOU2011480}. The estimator gives an initial guess of $E_{max}$, and we set $E_{min}$ as negative $E_{max}$. For this setting we have utilized the symmetry of the DOS, a bell-shape profile centered at zero, in the many-spin systems. It is important that $E_{max}$ and $E_{min}$ must bound all eigenvalues both from above and below. Otherwise, eigenstates with the largest absolute eigenvalues may also be magnified through the Delta filtering, which leads to the failure of the DELDAV method. For the interior states, we directly input the previously found $E_{max}$ and $E_{min}$ since typically they are quite easy to be searched for.

We note that the DELDAV method inherits the remarkable flexibility of the Davidson-type methods , which is beneficial to save memory and reduce orthogonalization cost. Therefore, we employ both the inner-outer restart technique, which relaxes the requirement for memory, and the block filter technique, which means several states are filtered during a single iteration, in programming of the DELDAV method~\cite{cd2}. This approach has been proved to converge rapidly with $d=k_{want}+c$, where $c$ is a positive integer, in the tests of the CD method~\cite{cd2}. Such a requirement of the subspace dimension is different from that of the implicitly restarted Arnoldi method (ARPACK), which needs $d\approx 2k_{want}$ to compute $k_{want}$ eigenpairs efficiently~\cite{Arpack}.

The cutoff expansion order $K$ depends on the interval length $L$ of wanted eigenvalues, which is composed of the number of required eigenstates $k_{want}$ and the local DOS $\rho$ of the system around $\lambda$. Roughly speaking, $K \sim L^{-1} = \rho \times k_{want}^{-1}$. The approximate distribution of DOS may be efficiently calculated through the Fourier transformation of a time evolved wave function or through a better estimation method given in~\cite{dosestimation}.

The convergence criterion is the residual norm $||r||=\sqrt{\left<\psi\left| (\mathcal{H}-\left<\mathcal{H}\right>)^2\right|\psi\right>}$ being less than a required numerical error $\epsilon$, where $\left<\mathcal{H}\right>=\left<\psi\left| \mathcal{H}\right| \psi\right>$ is the average energy of the state $\left|\psi\right>$ and for a converged state $\left<\mathcal{H}\right>$ represents the computed eigenvalue. It has been shown that $||r||$ gives an upper bound on the true error (absolute error) of the eigenvalue~\cite{YSaad}.

\section{Numerical results}
\label{sec:result}

We apply the DELDAV method to the eigenvalue problem in quantum spin-1/2 systems containing two-body interactions. Such systems are good model for investigating a large class of important problems in quantum computing, solid state theory, and quantum statistics~\cite{QCQI, QPT, QPTT}. Exact eigenvalues and eigenstates help us in understanding the physics behind the complicated model and often serve as a benchmark to evaluate other approximate methods as well.

In general, the system consists of $N$ spins and $M$ pairs of coupling, where $M$ ranges from $1$ to $(N^2-N)/2$. The Hamiltonian is given by
\begin{equation}
\mathcal{H}=\sum_{i<j}{\sum_{\alpha =x,y,z}{J_{ij}^{\alpha}S_{i}^{\alpha}S_{j}^{\alpha}}}+\sum_i{\sum_{\alpha =x,y,z}{H_{i}^{\alpha}S_{i}^{\alpha}}} \quad ,
\label{eq:ham}
\end{equation}
where the first term is the spin interaction with $J_{ij}^{\alpha}$ being interaction constants and the second term is the action of local magnetic fields $\textbf{H}_i$.

We specify the above many-spin model by two typical physical systems. One is the disordered one-dimensional transverse field Ising model~\cite{PhysRevB.51.6411}, where the Hamiltonian is
\begin{equation}
\mathcal{H}=\sum_{i=1}^{N-1}J_{i,i+1}\sigma_i^x \sigma_{i+1}^x+\sum_{i=1}^N \Gamma_i^z \sigma_i^z,
\end{equation}
with $\sigma_i$ the Pauli matrices for the spin $i$. This system is exactly solvable by Jordan-Wigner transformation~\cite{QPT}, making it an ideal correctness checker for the DELDAV method. The nearest neighbor exchange interaction constants $J_{i,i+1}$ are random numbers that uniformly distributed in $[-J/\sqrt{N},J/\sqrt{N}]$ with $J=10$. The local random magnetic fields are represented by $\Gamma^z_i$, which are random numbers that uniformly distributed in the interval $[0,\Gamma]$ with $\Gamma=1$.

Another system is the spin glass shards~\cite{sgs}, which represents a class of global-range interacting systems that require relatively large bond dimensions to be tackled by the DMRG methods~\cite{RevModPhys.77.259}. The Hamiltonian describing the system is
\begin{equation}
\mathcal{H}=\sum_{ i<j}J_{ij}\sigma_i^x\sigma_{j}^x+\sum_i \Gamma_i^z\sigma_i^z.
\end{equation}
All symbols and parameters are the same as that of the above Ising model, except that the first summation runs over all possible spin pairs. This system is interesting because it presents two crossovers from integrability to quantum chaos and back to integrability again. In the limit $J / \Gamma \rightarrow 0$, the ground state is paramagnetic with all spins in the local field direction and the system is integrable~\cite{sgs}. In the opposite limit $J/\Gamma\rightarrow \infty$, the ground state is spin glass and the system is also integrable since there are $N$ operators ($\sigma_i^x$) commuting with the Hamiltonian. A quantum chaos region exists between these two limits. $J=10 \,\Gamma$ is approximately the border from the quantum chaos to the integrable (the spin glass side) when $N=20$~\cite{sgs}.

Despite the asymptotic advantages of the Chebyshev expansion and the small band width of the Delta filter, it is not a priori clear if the DELDAV method is efficient for the real physical systems. We perform numerical tests to present the correctness, efficiency, and numerical robustness of the DELDAV method. All the timing information reported in this paper is obtained from calculations on the Intel(R) Xeon(R) CPU E5-2680 v3, using sequential mode. The convergence criterion is set as $\lVert r \rVert<\epsilon$ with $\epsilon =10^{-10}$.

We set $k_{want}=10$, i.e., to find out the $10$ eigenstates with eigenvalues closest to a given $\lambda$. In fact, we are able to calculate hundreds of eigenvalues with limited extra time consumption. For example, for the 15-spin Ising model, we are able to calculate $200$ eigenpairs in $5278$ CPU seconds versus $10$ eigenpairs in $1408$ CPU seconds. But for simplicity we consider the computation of $10$ eigenpairs henceforth. The ground cluster indicates the lowest $10$ eigenpairs while the central cluster indicates the $10$ eigenpairs closest to $\lambda=0$. 

\begin{figure}
\includegraphics[width=3.25 in]{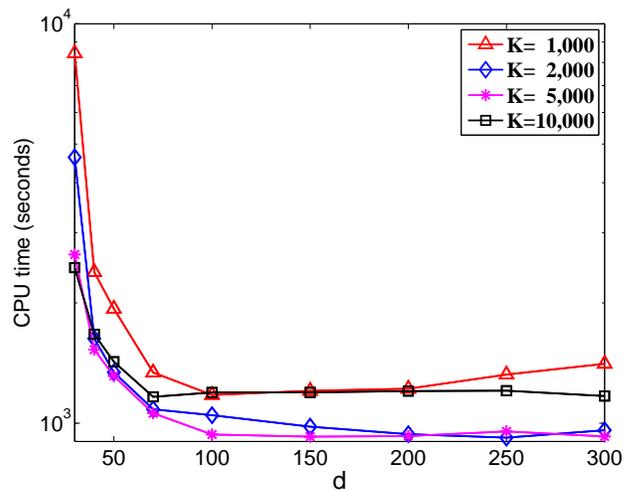}
\caption{\label{fig:ord} (Color online.) The convergence time of the DELDAV method versus the subspace dimension $d$,  with the cut off orders $K=1,000$ (red line with triangles), $2,000$ (blue line with diamonds), $5,000$ (pink line with asterisks), and $10,000$ (black line with squares). The tests are performed for calculating the central cluster in a $15$-spin Ising model. The robustness of the DELDAV method is revealed, as the convergence time is almost independent of $d$ and $K$ in a certain wide region.}
\end{figure}

The cutoff order of Chebyshev expansion is defined as $K= \lfloor 0.59\rho \rfloor$, where $0.59$ is determined according to the performance of our method for the $15$-spin Ising model. The subspace dimension $d=50$ in all the calculations except for the tests shown in Fig.~\ref{fig:ord}. The initial trial states in every test is a block of $3$ random pure states. For each numerical test, results of finding a certain $10$-eigenstate cluster are obtained. The computation of a 10-eigenstate cluster for Hilbert space dimension $2^{20}\sim 10^6$ requires around $1.2$ GB of memory.

\begin{figure}
\includegraphics[width=3.25 in]{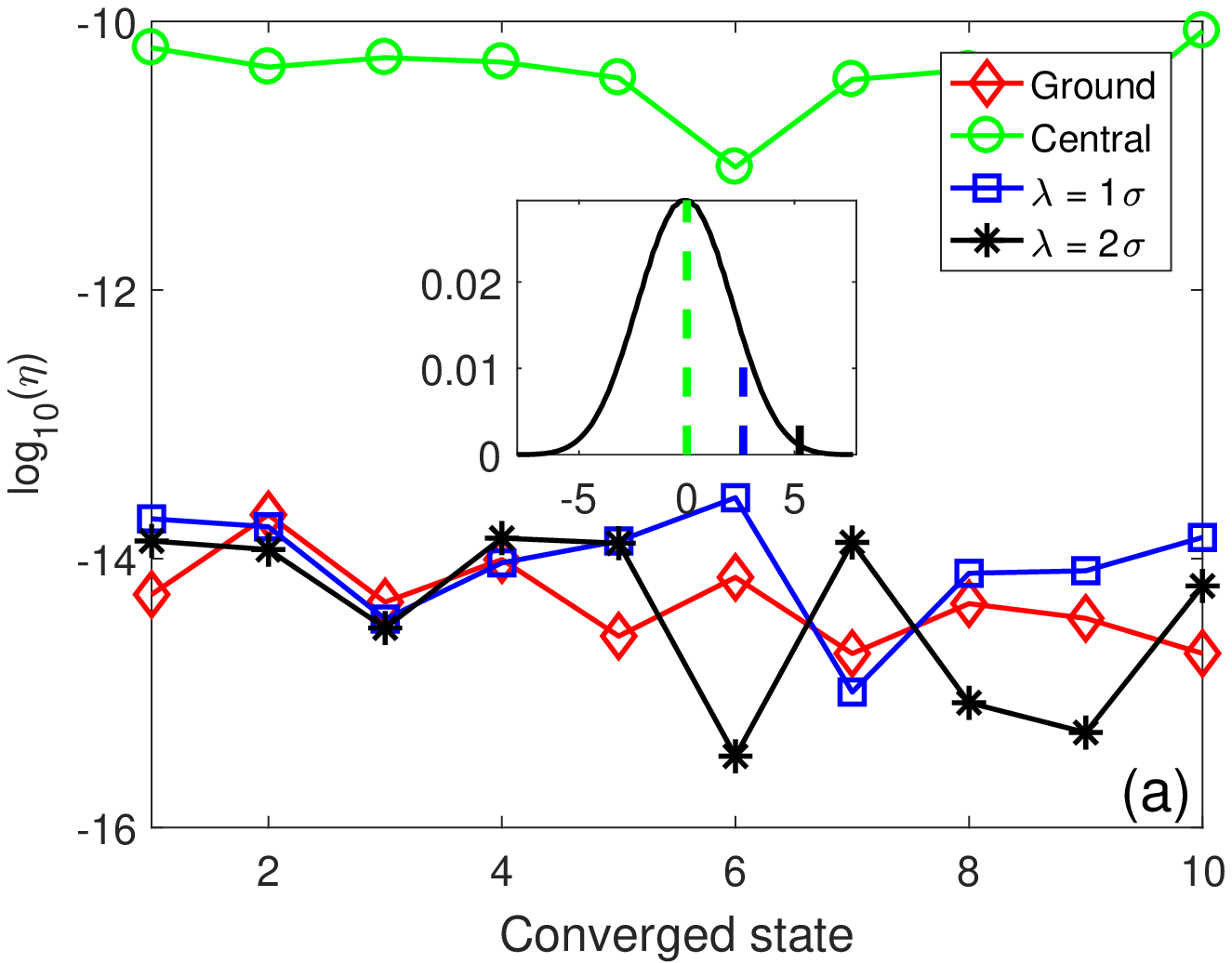}
\\
\includegraphics[width=3.25 in]{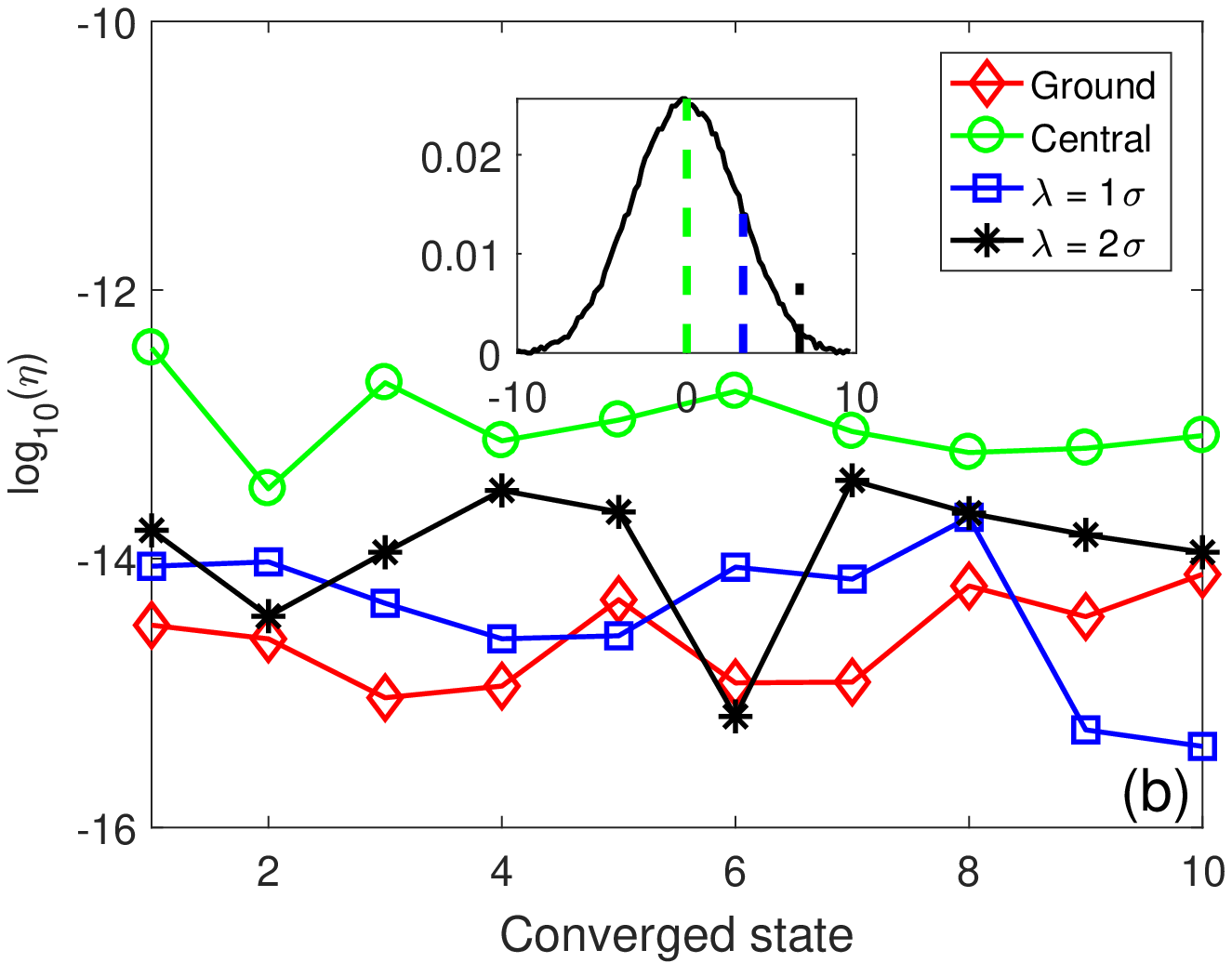}
\caption{\label{fig:accuracy} (Color online.) The relative error of the $10$ converged states in ground clusters (red lines with diamonds), central clusters (green lines with circles), the $1\sigma$ clusters (blue lines with squares), and the $2\sigma$ clusters (black lines with asterisks), is shown for (a) Ising model with $N=20$ and (b) spin glass shards model with $N=13$. The ground clusters are ordered by eigenvalues while the others by the distance between $\lambda$ and eigenvalues, in an increasing order. Insets present the normalized DOS for the systems and the location of different $\lambda$'s (vertical dashed lines). The $x$-axis of the insets is the system energy measured by $\Gamma$, and the $y$-axis the normalized DOS.}
\end{figure}

We present in Fig.~\ref{fig:accuracy} the relative error $\eta=|(E-E_{exact})/E_{exact}|$ of the computed eigenvalues for the Ising model with $N=20$ and the spin glass shards with $N=13$, where $E$ indicates the certain eigenvalue computed by the DELDAV method. The $x$-axis is the index of the converged states in the $10$-eigenstate cluster. Full exact eigenvalues of both systems have been obtained by other reliable methods. For the Ising model, the Jordan-Wigner transformation is used, while the spin glass shards is fully diagonalized through the subroutine ZHEEVR in LAPACK~\cite{Anderson1999LAPACK}. For each system, four $10$-eigenstate clusters located at the region with different local DOSs are calculated by our method, as shown in the insets in Fig.~\ref{fig:accuracy}. For the three interior clusters, $\lambda$ is chosen as $0$, $1\sigma$, and $2\sigma$, where $\sigma$ is the standard deviation of the Gaussian-like distribution of the DOS.

Clearly, our results for both systems agree well with the exact results at different $\lambda$'s with various local DOSs, including the extreme and the interior states (especially the central states). Note that although the absolute errors converge to a similar level, the relative errors $\eta$ for the central states show different behavior. The relatively large $\eta$ for the central states is due to the smallness of $E_{exact}$, which becomes extremely tiny for large $N$. In fact, for the Ising model with $20$ spins, the exact eigenvalues of the Hamiltonian $\mathcal{H}$ at the central region are about $10^{-5}\Gamma$, while for the spin glass shards with $13$ spins are about $10^{-3}\Gamma$.

\begin{figure}
\includegraphics[width=3.25 in]{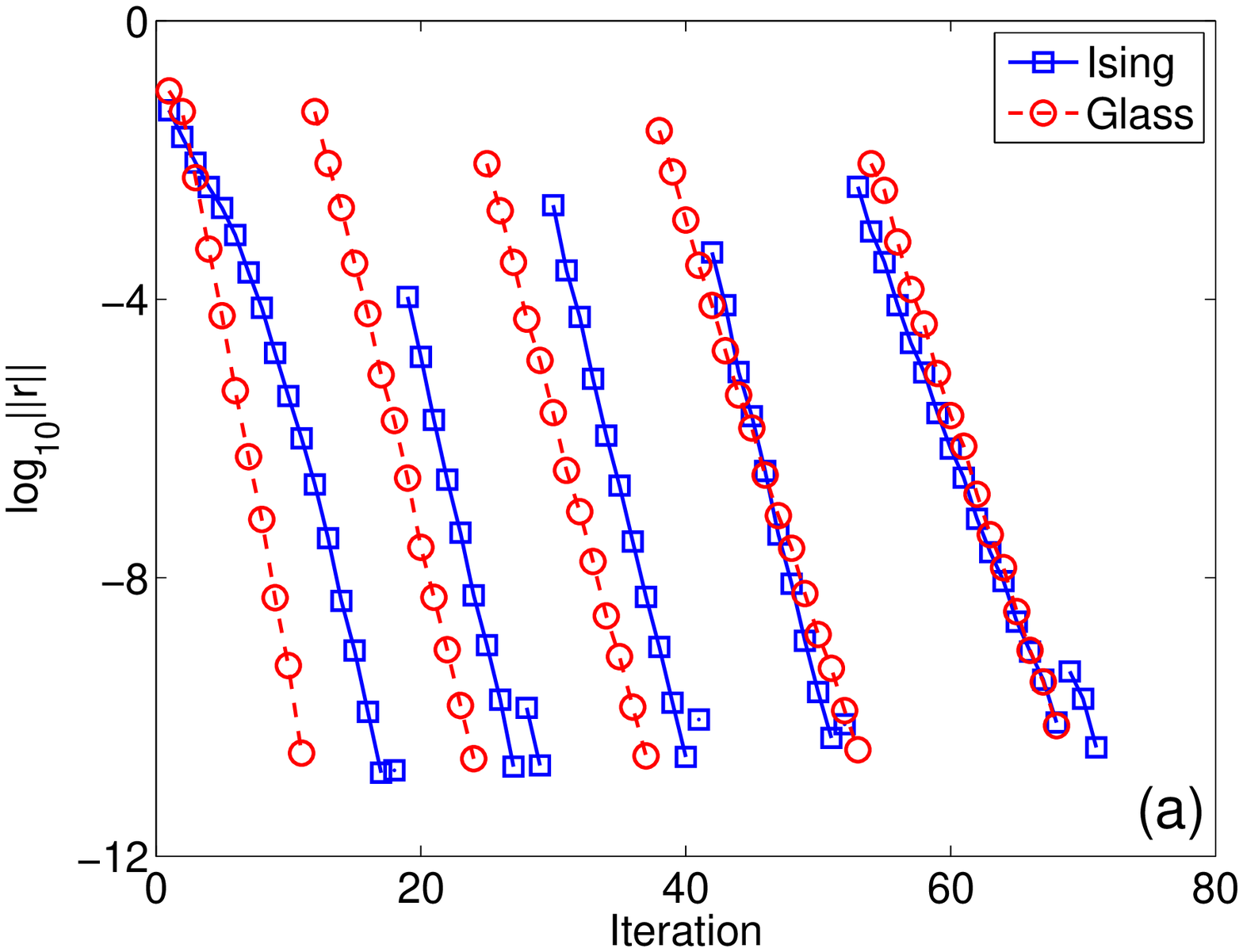}
\\
\includegraphics[width=3.25 in]{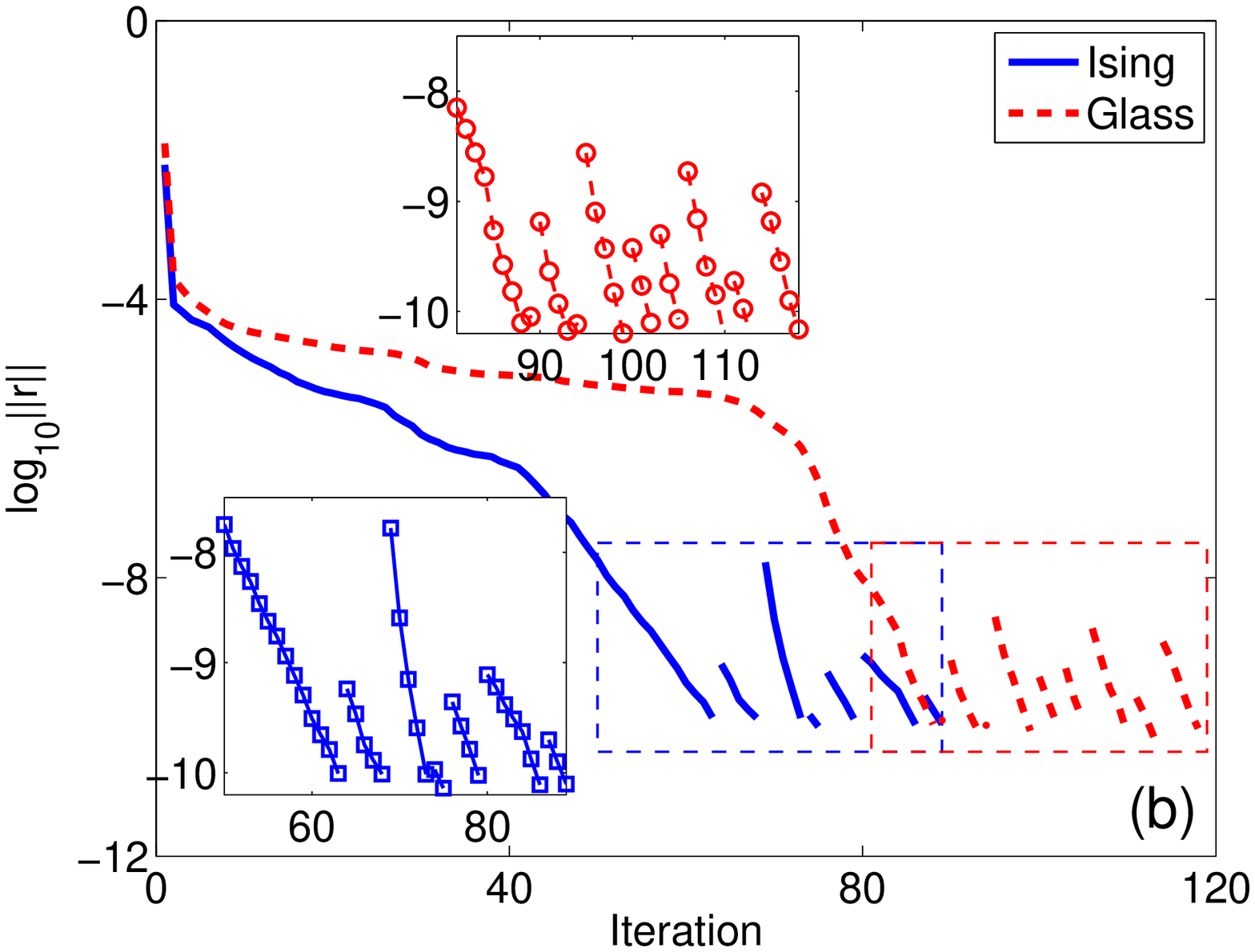}
\caption{\label{fig:err} (Color online.) Convergence processes measured by the residual norm for the computation of a $10$-eigenstate cluster in Ising model (blue solid lines with squares) and spin glass shards (red dashed lines with circles). (a) Chebyshev filtering process for ground clusters, both systems are of size $N=25$; (b) Delta filtering process for central clusters, with system size $N=20$ for Ising model and $N=19$ for spin glass shards. Each line corresponds to the convergence process of an eigenstate or a pair of close eigenstates (near-degenerate), since close eigenstates may converge simultaneously.}
\end{figure}

We present in Fig.~\ref{fig:err} the convergence processes of the CD method for the ground clusters and the DELDAV method for the central clusters. The $x$-axis shows the number of iterations, and during each iteration there is a filtration either by the Chebyshev filter or by the Delta filter, followed by a subspace diagonalization. The iterative filtering implies an effectively growing $k$ in Eq.~(\ref{eq:chevo}) for both filters. The $y$-axis represents the energy error bound for the state which is currently the closest to converge. We note that the eigenstates situated closer to $\lambda$ ($\lambda\le E_{min}$ for calculating the ground clusters) would show a faster convergence rate, thus the various convergence lines are ordered by the distance between the eigenvalues and $\lambda$. One may view the convergence process as a special dynamical evolution that concentrates the initial random wave function, which distributes widely in energy basis, into a sharp wave packet around $\lambda$.

We set $K=50$ for the Chebyshev filter in both systems. The straight lines in Fig.~\ref{fig:err}(a) confirm the exponential damping by the Chebyshev filtering, as also shown in Eq.~(\ref{eq:chevo}). However, after the fast convergence of the first two near-degenerate states, the residual norm jump to a rather high value~\cite{cd1}. This sudden change might be due to the over-damping of the unconverged states. For the small components, the numerical error will run in and be amplified after the orthonormalization. One thus needs to amplify the components of the nearest to converge states again through Chebyshev filtering in the following iterations.

For computation of the central clusters, we employ the DELDAV method for both spin systems since the CD method is not capable to search these eigenvalues. For the Delta filters, we set $K=80,000$ in the Ising model and $K=50,000$ in the spin glass shards. The huge increase of $K$, compared to the Chebyshev filters, comes from the high local DOS in the central region (see the insets in Fig.~\ref{fig:accuracy}). In general, a high local DOS means that more states are squeezed in an energy bin (near-degenerate) and the average level spacing is small. In order to discriminate these states, one needs a much smaller bandwidth filter, thus a very large $K$.

The convergence processes for the central clusters are shown in Fig.~\ref{fig:err}(b). Those two curves clearly show three stages. The first stage exhibits a fast decay of the residual norm, because the components of the eigenstates with eigenvalues far from $\lambda$ diminish quickly after the Delta filtering, as shown in Fig.~\ref{fig:shape}(d-f). The second one is, however, a plateau. We suppose the slowdown of the convergence inside the plateau might originates from the flatness of the Delta filters nearby $\lambda$. Without the subspace diagonalization, the plateau would persist for a rather long time, as further confirmed in Appendix~\ref{sec:appen}. It is interesting that the third stage shows an exponential decay of $\lVert r \rVert$ (see the two insets), which is similar to the behavior in Fig.~\ref{fig:err}(a). Obviously, the fast decay in this stage is due to the combined effect of the Delta filtering and the subspace diagonalization, which is the essential idea of the Davidson-type methods.

\begin{figure}
\includegraphics[width=3.25 in]{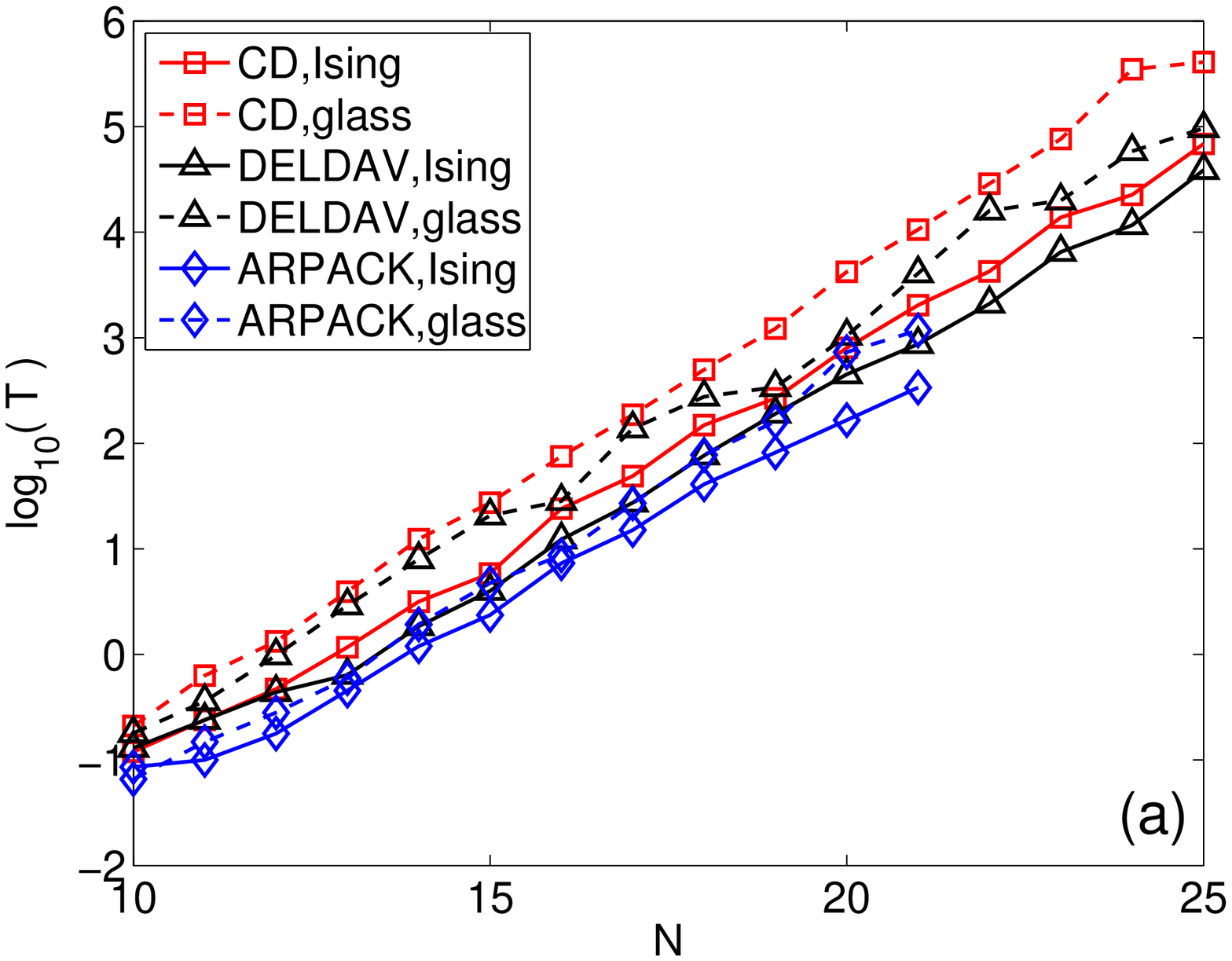}

\includegraphics[width=3.25 in]{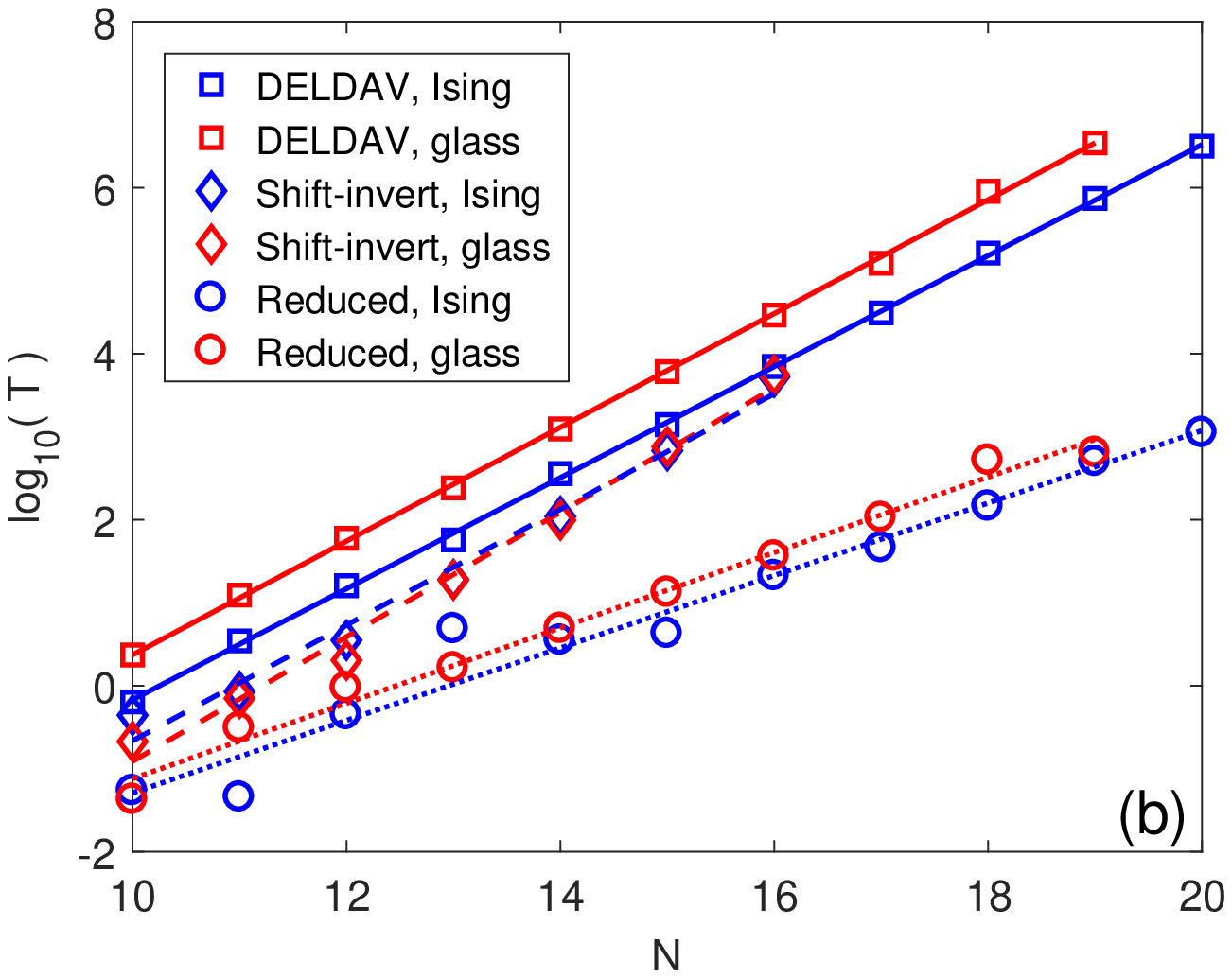}
\caption{\label{fig:scaling} (Color online.) Scaling behavior measured by the convergence time $T$ (CPU seconds) versus system size $N$. (a) Comparison of the CD method (red lines with squares), the DELDAV method (black lines with triangles), and ARPACK (blue lines with diamonds) for calculating the ground clusters of the Ising model (solid lines) and of the spin glass shards (dashed lines). All of them share a similar exponential growth with $N$. (b) Scaling lines of calculating the central clusters, with the DELDAV method (solid lines with squares), the shift-invert method (dashed lines with diamonds) and the reduced scaling lines of the DELDAV method (dotted lines with circles) for the Ising model (blue) and the spin glass shards (red). All these lines are obtained via linear fitting. The shift-invert method possesses the maximum slope. The reduced scaling lines are close to those of (a).}
\end{figure}

\begin{table}
	\caption{\label{tab:scaling} Scaling constants $\alpha$ by exponential fitting of the DELDAV curves in Fig.~\ref{fig:scaling}.}
	\begin{tabular}{c|c|c|c}
		\hline\hline
		$\alpha$ & Ground & Central, original & Central, reduced \\
		\hline
		Ising & 0.86 & 1.54 & 1.00 \\
		Glass & 0.90 & 1.58 & 1.05 \\
		\hline\hline
	\end{tabular}
\end{table}

The scaling behavior of the DELDAV method for finding the ground and central clusters of the Ising model and the spin glass shards are presented in Fig.~\ref{fig:scaling}. For the ground clusters, we also present scaling lines of the CD method and ARPACK for comparisons. ARPACK is one of the most robust and efficient eigensolvers in practice~\cite{implicit, Stewart}. It provides the matrix-vector products mode for extreme cluster calculations and often serves as a benchmark~\cite{Arpack}. We define the convergence time $T$ as the CPU time (seconds) for calculating a $10$-eigenstate cluster. As shown in Fig.~\ref{fig:scaling}(a), the scaling of runtime versus system size between these methods is comparable. In detail, ARPACK costs the least CPU time to converge, followed by the DELDAV method while the CD method performs the worst. All three methods have similar memory consumption since only the matrix-vector products are required.

We note, however, that both the CD method and ARPACK are not  suitable for solving interior eigenstates. As mentioned before, the CD method cannot directly find out the interior eigenstates. When combined with the transformation $(\mathcal{H}-\lambda)^2$, the tiny level spacings ($10^{-7}$ for $\mathcal{G}$) at the central region are squared ($10^{-14}$ for $\mathcal{G}^2$). Roughly speaking, it means $10^7$ times longer for the convergence, which is an unacceptable long time. In finding the interior eigenstates, ARPACK provides only the shift-invert mode instead of the matrix-vector products mode. This mode is essentially the shift-invert method, i.e., replacing the original Hamiltonian $\mathcal{H}$ by its inverse $\mathcal{H}^{-1}$. The shift-invert method is widely used in calculating eigenpairs at the middle of the spectrum, to name a few, like in~\cite{scipost, PhysRevB.91.081103, PhysRevB.101.104201, PhysRevA.101.063617}. However, for large spin systems, the matrix sizes are extremely huge and one will quickly run out of memory when calculating the inverse of the Hamiltonian. On the contrary, the DELDAV method is applicable on the full spectrum while needing only matrix-vector products, which does not require an explicit Hamiltonian matrix representation. 

For the central clusters, we present in Fig.~\ref{fig:scaling}(b) the scaling lines of the DELDAV method for both the Ising model and the spin glass shards. Similarly, the DELDAV method exhibits an exponential scaling. However, the scaling line slopes of the central clusters are far higher compared to that of Fig.~\ref{fig:scaling}(a).  To compare quantitatively, we extract the scaling constants $\alpha$ by fitting the numerical results, where $\alpha$ is defined by $T = T_0 \exp(\alpha N)$. The values of $\alpha$ are shown in Table~\ref{tab:scaling} for the ground and central clusters. One may observe that $\alpha$ in Fig.~\ref{fig:scaling}(b) are indeed apparently larger than those in Fig.~\ref{fig:scaling}(a).

Since the DELDAV method works unbiased on the full spectrum, and the convergence of the eigenstates at larger local DOS regions requires a smaller bandwidth, we speculate that the larger $\alpha$ may be due to the larger local DOS $\rho$ in the central region of the spectrum. We thus define a reduced convergence time $T'=T\rho^{G}/\rho^{C}= T E_{b}^{C}/E_{b}^{G}$, with $\rho^{G,C}=10/E_b^{G,C}$ the local DOS of the found ground or central clusters, $E_b^{G,C}$ the energy band of the found ground or central clusters. The reduced convergence time eliminates the influence from various local DOSs. Not surprisingly, the asymptotic scaling of the reduced time almost overlaps for both spin systems and is close to that for the ground clusters in Fig.~\ref{fig:scaling}(a). As presented in Table~\ref{tab:scaling}, the reduced scaling constants, resulting from the reduced convergence time, are pretty close between the ground and the central clusters. The universal value of the reduced scaling constants could be a manifestation of the proportional relation between the convergence time and the local DOS.

In Fig.~\ref{fig:scaling}(b) we also present the scaling lines of the shift-invert technique combined with ARPACK in computing the central clusters for comparison. The extracted scaling constants of the shift-invert method are $\alpha=1.61$ for the Ising model and $\alpha=1.73$ for the spin glass shards, which are slightly higher than the DELDAV method. We note, however, that the LU decomposition (factorization) dominates the computation time for the shift-invert method when the system size $N\ge18$~\cite{scipost}. Thus for large systems its convergence time is approximately proportional to the cube of matrix size~\cite{ITNMN}, giving a scaling constant $\alpha\approx2.08$. Considering the similar convergence time of both methods for $N=16$ systems, it is evident that the DELDAV method converges faster for large enough spin systems. In addition, the shift-invert method requires 98 GB memory for $N=16$ systems. Such a requirement is much larger than that of 89 MB memory by the DELDAV method for the same system. Restricted by the memory capacity, with the shift-invert method we cannot perform calculations for system size $N\ge17$.  These comparisons clearly indicate the huge advantages of the DELDAV method over the shift-invert method for large spin systems.


\begin{figure}
\includegraphics[width=3.25 in]{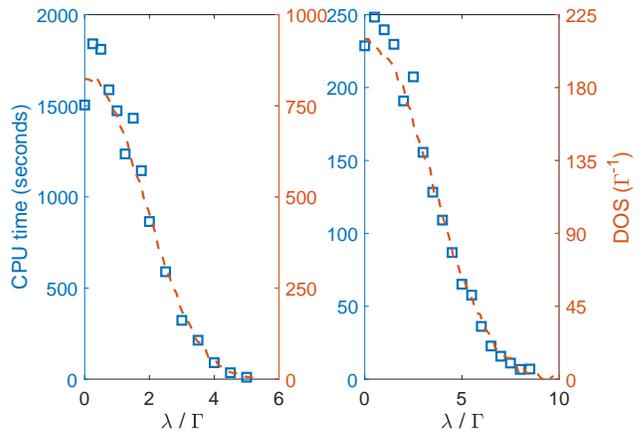}
\caption{\label{fig:dos} (Color online.) The convergence time (blue squares) of the DELDAV method for computing a variety of eigenvalue clusters and the local DOS (green dashed lines) for a $15$-spin Ising model (left) and for a $13$-spin glass shards (right). Obviously, the convergence time is proportional to the local DOS.}
\end{figure}

We present in Fig.~\ref{fig:dos} both the convergence time of the DELDAV method and the local DOS for the Ising model and the spin glass shards. We have performed two tests with various $\lambda$'s, the first for the Ising model with 15 spins while the second for the spin glass shards with 13 spins. The DOS for both systems are statistically constructed by their entire exact eigenvalues. Clearly, Fig.~\ref{fig:dos} shows that the DELDAV method is much more efficient in finding the interior states on the edge of the density profile than the central ones. It also implies a proportional relationship between the local DOS and the convergence time. This relation suggests the DELDAV method indeed has similar efficiency in finding the interior states (including the central ones) and the extreme states, in the sense that the local DOS being the same.

Besides the DOS, many other factors in practice may have influence on the convergence time and efficiency of the DELDAV method. Two factors, the subspace dimension $d$ and the cutoff order of Chebyshev expansion $K$, are controllable in the numerical calculations. In Fig.~\ref{fig:ord}, we present the dependence of the convergence time on $d$ and $K$.  The tests are performed for finding the central cluster in the 15-spin Ising model, with all parameters the same except for $d$ and $K$. In the small $d$ limit, we notice that a larger $K$ greatly speeds up the convergence of the program. While in the large $d$ limit, the convergence time is nearly independent of $d$ and $K$, indicating the robustness of the DELDAV method. This also suggests one to choose $d=100$ when 10 eigenvalues are wanted, in order to converge quickly and to save memory at same time.

\section{CONCLUSION}
\label{sec:con}
We propose the Delta-Davidson method to solve both the extreme and interior eigenvalue problems. The method hybridizes the advantages of the Delta filtering and the subspace diagonalization, which efficiently constructs the special subspace and solves the highly near-degenerate problem. We present an explicit algorithm and the specific details to apply the DELDAV method to quantum spin models. The numerical experiments for the Ising model and the spin glass shards confirm the exactness, efficiency and robustness of the DELDAV method. Although all three methods share roughly the same scaling behavior for extreme eigenvalue calculations, the DELDAV method has an obvious advantage over the Chebyshev-Davidson method and Arnoldi method with its capability for finding the interior eigenstates. As for the central eigenstates, the DELDAV method shows a potentially faster convergence rate for a large enough system and consumes far less memory, compared to the shift-invert method. For the interior regions with small density of the states, the DELDAV method may converge several ten times faster. The small requirement of memory makes it possible to simultaneously run several threads to compute different disorder settings in many body localization problems. We believe these features render the DELDAV method a competitive instrument for eigenpairs calculations.  

\begin{figure}
\includegraphics[width=3.25 in]{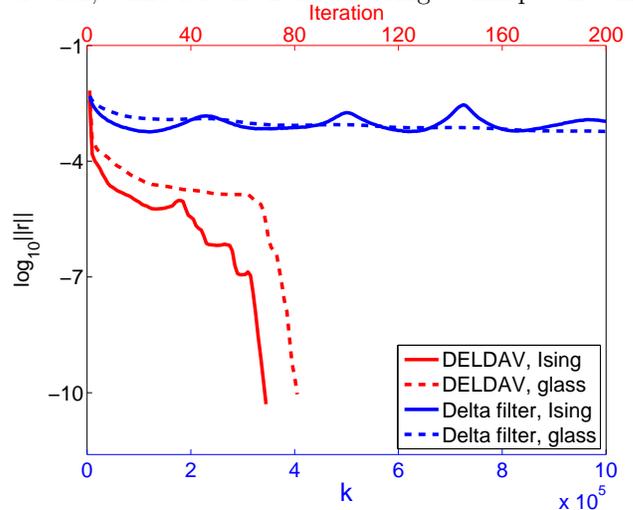}
\caption{\label{fig:cmp} (Color online.) Comparison of the convergence processes for the DELDAV method (top axis, red lines) and the Delta filtering (bottom axis, blue lines) in computing a single eigenvalue at the largest DOS region for the Ising system (solid lines) and the spin glass shards (dashed lines). Both systems are of size $N=15$. Clearly, the DELDAV method converges faster.}
\end{figure}

\appendix
\section{DELDAV vs Delta filtering}
\label{sec:appen}

To illustrate the advantages of the DELDAV method in dealing with the near-degenerate problem, we plot in Fig.~\ref{fig:cmp} the convergence processes in finding a single eigenvalue at the largest DOS region for both the DELDAV method and the Delta filtering technique. The specific eigenvalue to compute is the closest one to $\lambda$. The tests are performed for the two 15-spin systems, the Ising model with $\lambda=10^{-4}$ and the spin glass shards with $\lambda=0$. The level spacings of the rescaled operator $\mathcal{G}$ for both systems are about $10^{-5}$. For both methods the initial states are random. For the DELDAV method, the subspace dimension $d=50$ and the cutoff order $K=5,000$, i.e., the Delta filter is $\delta_{5,000}(\mathcal{H}-\lambda)$ in each iteration; while for the Delta filtering technique the expansion order $k$ keeps growing and we record the residual norm $||r||$ when $k$ is increased by $5,000$. 

The convergence lines for the two methods clearly show a different convergence rate. Although the DELDAV method presents a rather slow convergence during the plateau stage, it quickly speeds up to a fast convergence rate after accumulation of the good base states. On the contrary, the Delta filtering technique alone keeps the slow convergence for a rather long time. The advantages of the DELDAV method over the Delta filtering technique is thus confirmed.

\begin{acknowledgments}
We thank A.-Q. Shi for discussions. This work is supported by the NSFC Grant No. 91836101, No. 11574239, and No. U1930201.
\end{acknowledgments}

\bibliography{Deldav}

\providecommand{\noopsort}[1]{}\providecommand{\singleletter}[1]{#1}%
\begin{thebibliography}{57}%
\makeatletter
\providecommand \@ifxundefined [1]{%
 \@ifx{#1\undefined}
}%
\providecommand \@ifnum [1]{%
 \ifnum #1\expandafter \@firstoftwo
 \else \expandafter \@secondoftwo
 \fi
}%
\providecommand \@ifx [1]{%
 \ifx #1\expandafter \@firstoftwo
 \else \expandafter \@secondoftwo
 \fi
}%
\providecommand \natexlab [1]{#1}%
\providecommand \enquote  [1]{``#1''}%
\providecommand \bibnamefont  [1]{#1}%
\providecommand \bibfnamefont [1]{#1}%
\providecommand \citenamefont [1]{#1}%
\providecommand \href@noop [0]{\@secondoftwo}%
\providecommand \href [0]{\begingroup \@sanitize@url \@href}%
\providecommand \@href[1]{\@@startlink{#1}\@@href}%
\providecommand \@@href[1]{\endgroup#1\@@endlink}%
\providecommand \@sanitize@url [0]{\catcode `\\12\catcode `\$12\catcode
  `\&12\catcode `\#12\catcode `\^12\catcode `\_12\catcode `\%12\relax}%
\providecommand \@@startlink[1]{}%
\providecommand \@@endlink[0]{}%
\providecommand \url  [0]{\begingroup\@sanitize@url \@url }%
\providecommand \@url [1]{\endgroup\@href {#1}{\urlprefix }}%
\providecommand \urlprefix  [0]{URL }%
\providecommand \Eprint [0]{\href }%
\providecommand \doibase [0]{http://dx.doi.org/}%
\providecommand \selectlanguage [0]{\@gobble}%
\providecommand \bibinfo  [0]{\@secondoftwo}%
\providecommand \bibfield  [0]{\@secondoftwo}%
\providecommand \translation [1]{[#1]}%
\providecommand \BibitemOpen [0]{}%
\providecommand \bibitemStop [0]{}%
\providecommand \bibitemNoStop [0]{.\EOS\space}%
\providecommand \EOS [0]{\spacefactor3000\relax}%
\providecommand \BibitemShut  [1]{\csname bibitem#1\endcsname}%
\let\auto@bib@innerbib\@empty
\bibitem [{\citenamefont {Nielsen}\ and\ \citenamefont {Chuang}(2011)}]{QCQI}%
  \BibitemOpen
  \bibfield  {author} {\bibinfo {author} {\bibfnamefont {M.~A.}\ \bibnamefont
  {Nielsen}}\ and\ \bibinfo {author} {\bibfnamefont {I.~L.}\ \bibnamefont
  {Chuang}},\ }\href@noop {} {\emph {\bibinfo {title} {Quantum Computation and
  Quantum Information}}}\ (\bibinfo  {publisher} {Cambridge University Press,
  Cambridge, England},\ \bibinfo {year} {2011})\BibitemShut {NoStop}%
\bibitem [{\citenamefont {Sachdev}(2011)}]{QPT}%
  \BibitemOpen
  \bibfield  {author} {\bibinfo {author} {\bibfnamefont {S.}~\bibnamefont
  {Sachdev}},\ }\href@noop {} {\emph {\bibinfo {title} {Quantum Phase
  Transitions}}}\ (\bibinfo  {publisher} {Cambridge University Press,
  Cambridge, England},\ \bibinfo {year} {2011})\BibitemShut {NoStop}%
\bibitem [{\citenamefont {Bogolubov}\ and\ \citenamefont
  {N.~N.~Bogolubov}(2009)}]{IQSM}%
  \BibitemOpen
  \bibfield  {author} {\bibinfo {author} {\bibfnamefont {N.~N.}\ \bibnamefont
  {Bogolubov}}\ and\ \bibinfo {author} {\bibfnamefont {J.}~\bibnamefont
  {N.~N.~Bogolubov}},\ }\href@noop {} {\emph {\bibinfo {title} {Introduction to
  Quantum Statistical Mechanics}}}\ (\bibinfo  {publisher} {World Scientific
  Publishing Co Pte Ltd, Singapore},\ \bibinfo {year} {2009})\BibitemShut
  {NoStop}%
\bibitem [{\citenamefont {Dutta}\ \emph {et~al.}(2015)\citenamefont {Dutta},
  \citenamefont {Aeppli}, \citenamefont {Chakrabarti}, \citenamefont
  {Divakaran}, \citenamefont {Rosenbaum},\ and\ \citenamefont {Sen}}]{QPTT}%
  \BibitemOpen
  \bibfield  {author} {\bibinfo {author} {\bibfnamefont {A.}~\bibnamefont
  {Dutta}}, \bibinfo {author} {\bibfnamefont {G.}~\bibnamefont {Aeppli}},
  \bibinfo {author} {\bibfnamefont {B.~K.}\ \bibnamefont {Chakrabarti}},
  \bibinfo {author} {\bibfnamefont {U.}~\bibnamefont {Divakaran}}, \bibinfo
  {author} {\bibfnamefont {T.~F.}\ \bibnamefont {Rosenbaum}}, \ and\ \bibinfo
  {author} {\bibfnamefont {D.}~\bibnamefont {Sen}},\ }\href@noop {} {\emph
  {\bibinfo {title} {Quantum Phase Transitions in Transverse Field Spin Models:
  From Statistical Physics to Quantum Information}}}\ (\bibinfo  {publisher}
  {Cambridge University Press, Cambridge, England},\ \bibinfo {year}
  {2015})\BibitemShut {NoStop}%
\bibitem [{\citenamefont {Bortz}\ \emph {et~al.}(2010)\citenamefont {Bortz},
  \citenamefont {Eggert}, \citenamefont {Schneider}, \citenamefont
  {St\"ubner},\ and\ \citenamefont {Stolze}}]{PhysRevB.82.161308}%
  \BibitemOpen
  \bibfield  {author} {\bibinfo {author} {\bibfnamefont {M.}~\bibnamefont
  {Bortz}}, \bibinfo {author} {\bibfnamefont {S.}~\bibnamefont {Eggert}},
  \bibinfo {author} {\bibfnamefont {C.}~\bibnamefont {Schneider}}, \bibinfo
  {author} {\bibfnamefont {R.}~\bibnamefont {St\"ubner}}, \ and\ \bibinfo
  {author} {\bibfnamefont {J.}~\bibnamefont {Stolze}},\ }\href {\doibase
  10.1103/PhysRevB.82.161308} {\bibfield  {journal} {\bibinfo  {journal} {Phys.
  Rev. B}\ }\textbf {\bibinfo {volume} {82}},\ \bibinfo {pages} {161308(R)}
  (\bibinfo {year} {2010})}\BibitemShut {NoStop}%
\bibitem [{\citenamefont {Dobrovitski}\ and\ \citenamefont
  {De~Raedt}(2003)}]{Dob}%
  \BibitemOpen
  \bibfield  {author} {\bibinfo {author} {\bibfnamefont {V.~V.}\ \bibnamefont
  {Dobrovitski}}\ and\ \bibinfo {author} {\bibfnamefont {H.~A.}\ \bibnamefont
  {De~Raedt}},\ }\href {\doibase 10.1103/PhysRevE.67.056702} {\bibfield
  {journal} {\bibinfo  {journal} {Phys. Rev. E}\ }\textbf {\bibinfo {volume}
  {67}},\ \bibinfo {pages} {056702} (\bibinfo {year} {2003})}\BibitemShut
  {NoStop}%
\bibitem [{\citenamefont {Sandvik}(2010)}]{PhysRevLett.104.137204}%
  \BibitemOpen
  \bibfield  {author} {\bibinfo {author} {\bibfnamefont {A.~W.}\ \bibnamefont
  {Sandvik}},\ }\href {\doibase 10.1103/PhysRevLett.104.137204} {\bibfield
  {journal} {\bibinfo  {journal} {Phys. Rev. Lett.}\ }\textbf {\bibinfo
  {volume} {104}},\ \bibinfo {pages} {137204} (\bibinfo {year}
  {2010})}\BibitemShut {NoStop}%
\bibitem [{\citenamefont {Blankenbecler}\ and\ \citenamefont
  {Sugar}(1983)}]{PhysRevD.27.1304}%
  \BibitemOpen
  \bibfield  {author} {\bibinfo {author} {\bibfnamefont {R.}~\bibnamefont
  {Blankenbecler}}\ and\ \bibinfo {author} {\bibfnamefont {R.~L.}\ \bibnamefont
  {Sugar}},\ }\href {\doibase 10.1103/PhysRevD.27.1304} {\bibfield  {journal}
  {\bibinfo  {journal} {Phys. Rev. D}\ }\textbf {\bibinfo {volume} {27}},\
  \bibinfo {pages} {1304} (\bibinfo {year} {1983})}\BibitemShut {NoStop}%
\bibitem [{\citenamefont {Sandvik}\ and\ \citenamefont
  {Kurkij\"arvi}(1991)}]{PhysRevB.43.5950}%
  \BibitemOpen
  \bibfield  {author} {\bibinfo {author} {\bibfnamefont {A.~W.}\ \bibnamefont
  {Sandvik}}\ and\ \bibinfo {author} {\bibfnamefont {J.}~\bibnamefont
  {Kurkij\"arvi}},\ }\href {\doibase 10.1103/PhysRevB.43.5950} {\bibfield
  {journal} {\bibinfo  {journal} {Phys. Rev. B}\ }\textbf {\bibinfo {volume}
  {43}},\ \bibinfo {pages} {5950} (\bibinfo {year} {1991})}\BibitemShut
  {NoStop}%
\bibitem [{\citenamefont {White}(1992)}]{PhysRevLett.69.2863}%
  \BibitemOpen
  \bibfield  {author} {\bibinfo {author} {\bibfnamefont {S.~R.}\ \bibnamefont
  {White}},\ }\href {\doibase 10.1103/PhysRevLett.69.2863} {\bibfield
  {journal} {\bibinfo  {journal} {Phys. Rev. Lett.}\ }\textbf {\bibinfo
  {volume} {69}},\ \bibinfo {pages} {2863} (\bibinfo {year}
  {1992})}\BibitemShut {NoStop}%
\bibitem [{\citenamefont {Vidal}(2004)}]{PhysRevLett.93.040502}%
  \BibitemOpen
  \bibfield  {author} {\bibinfo {author} {\bibfnamefont {G.}~\bibnamefont
  {Vidal}},\ }\href {\doibase 10.1103/PhysRevLett.93.040502} {\bibfield
  {journal} {\bibinfo  {journal} {Phys. Rev. Lett.}\ }\textbf {\bibinfo
  {volume} {93}},\ \bibinfo {pages} {040502} (\bibinfo {year}
  {2004})}\BibitemShut {NoStop}%
\bibitem [{\citenamefont {Vidal}(2007)}]{PhysRevLett.99.220405}%
  \BibitemOpen
  \bibfield  {author} {\bibinfo {author} {\bibfnamefont {G.}~\bibnamefont
  {Vidal}},\ }\href {\doibase 10.1103/PhysRevLett.99.220405} {\bibfield
  {journal} {\bibinfo  {journal} {Phys. Rev. Lett.}\ }\textbf {\bibinfo
  {volume} {99}},\ \bibinfo {pages} {220405} (\bibinfo {year}
  {2007})}\BibitemShut {NoStop}%
\bibitem [{\citenamefont {Verstraete}\ \emph {et~al.}(2008)\citenamefont
  {Verstraete}, \citenamefont {Murg},\ and\ \citenamefont
  {Cirac}}]{Verstraete2008Matrix}%
  \BibitemOpen
  \bibfield  {author} {\bibinfo {author} {\bibfnamefont {F.}~\bibnamefont
  {Verstraete}}, \bibinfo {author} {\bibfnamefont {V.}~\bibnamefont {Murg}}, \
  and\ \bibinfo {author} {\bibfnamefont {J.}~\bibnamefont {Cirac}},\
  }\href@noop {} {\bibfield  {journal} {\bibinfo  {journal} {Adv. Phys.}\
  }\textbf {\bibinfo {volume} {57}},\ \bibinfo {pages} {143} (\bibinfo {year}
  {2008})}\BibitemShut {NoStop}%
\bibitem [{\citenamefont {Orús}(2014)}]{ORUS2014117}%
  \BibitemOpen
  \bibfield  {author} {\bibinfo {author} {\bibfnamefont {R.}~\bibnamefont
  {Orús}},\ }\href {\doibase https://doi.org/10.1016/j.aop.2014.06.013}
  {\bibfield  {journal} {\bibinfo  {journal} {Ann. Phys.}\ }\textbf {\bibinfo
  {volume} {349}},\ \bibinfo {pages} {117 } (\bibinfo {year}
  {2014})}\BibitemShut {NoStop}%
\bibitem [{\citenamefont {Eisert}\ \emph {et~al.}(2010)\citenamefont {Eisert},
  \citenamefont {Cramer},\ and\ \citenamefont {Plenio}}]{RevModPhys.82.277}%
  \BibitemOpen
  \bibfield  {author} {\bibinfo {author} {\bibfnamefont {J.}~\bibnamefont
  {Eisert}}, \bibinfo {author} {\bibfnamefont {M.}~\bibnamefont {Cramer}}, \
  and\ \bibinfo {author} {\bibfnamefont {M.~B.}\ \bibnamefont {Plenio}},\
  }\href {\doibase 10.1103/RevModPhys.82.277} {\bibfield  {journal} {\bibinfo
  {journal} {Rev. Mod. Phys.}\ }\textbf {\bibinfo {volume} {82}},\ \bibinfo
  {pages} {277} (\bibinfo {year} {2010})}\BibitemShut {NoStop}%
\bibitem [{\citenamefont {Loh}\ \emph {et~al.}(1990)\citenamefont {Loh},
  \citenamefont {Gubernatis}, \citenamefont {Scalettar}, \citenamefont {White},
  \citenamefont {Scalapino},\ and\ \citenamefont {Sugar}}]{PhysRevB.41.9301}%
  \BibitemOpen
  \bibfield  {author} {\bibinfo {author} {\bibfnamefont {E.~Y.}\ \bibnamefont
  {Loh}}, \bibinfo {author} {\bibfnamefont {J.~E.}\ \bibnamefont {Gubernatis}},
  \bibinfo {author} {\bibfnamefont {R.~T.}\ \bibnamefont {Scalettar}}, \bibinfo
  {author} {\bibfnamefont {S.~R.}\ \bibnamefont {White}}, \bibinfo {author}
  {\bibfnamefont {D.~J.}\ \bibnamefont {Scalapino}}, \ and\ \bibinfo {author}
  {\bibfnamefont {R.~L.}\ \bibnamefont {Sugar}},\ }\href {\doibase
  10.1103/PhysRevB.41.9301} {\bibfield  {journal} {\bibinfo  {journal} {Phys.
  Rev. B}\ }\textbf {\bibinfo {volume} {41}},\ \bibinfo {pages} {9301}
  (\bibinfo {year} {1990})}\BibitemShut {NoStop}%
\bibitem [{\citenamefont {Troyer}\ and\ \citenamefont
  {Wiese}(2005)}]{PhysRevLett.94.170201}%
  \BibitemOpen
  \bibfield  {author} {\bibinfo {author} {\bibfnamefont {M.}~\bibnamefont
  {Troyer}}\ and\ \bibinfo {author} {\bibfnamefont {U.-J.}\ \bibnamefont
  {Wiese}},\ }\href {\doibase 10.1103/PhysRevLett.94.170201} {\bibfield
  {journal} {\bibinfo  {journal} {Phys. Rev. Lett.}\ }\textbf {\bibinfo
  {volume} {94}},\ \bibinfo {pages} {170201} (\bibinfo {year}
  {2005})}\BibitemShut {NoStop}%
\bibitem [{\citenamefont {Lanczos}(1950)}]{Lanczos2018An}%
  \BibitemOpen
  \bibfield  {author} {\bibinfo {author} {\bibfnamefont {C.}~\bibnamefont
  {Lanczos}},\ }\href@noop {} {\bibfield  {journal} {\bibinfo  {journal} {J.
  Res. Natl. Bur. Stand.}\ }\textbf {\bibinfo {volume} {45}},\ \bibinfo {pages}
  {255} (\bibinfo {year} {1950})}\BibitemShut {NoStop}%
\bibitem [{\citenamefont {Vidmar}\ \emph {et~al.}(2018)\citenamefont {Vidmar},
  \citenamefont {Hackl}, \citenamefont {Bianchi},\ and\ \citenamefont
  {Rigol}}]{PhysRevLett.121.220602}%
  \BibitemOpen
  \bibfield  {author} {\bibinfo {author} {\bibfnamefont {L.}~\bibnamefont
  {Vidmar}}, \bibinfo {author} {\bibfnamefont {L.}~\bibnamefont {Hackl}},
  \bibinfo {author} {\bibfnamefont {E.}~\bibnamefont {Bianchi}}, \ and\
  \bibinfo {author} {\bibfnamefont {M.}~\bibnamefont {Rigol}},\ }\href
  {\doibase 10.1103/PhysRevLett.121.220602} {\bibfield  {journal} {\bibinfo
  {journal} {Phys. Rev. Lett.}\ }\textbf {\bibinfo {volume} {121}},\ \bibinfo
  {pages} {220602} (\bibinfo {year} {2018})}\BibitemShut {NoStop}%
\bibitem [{\citenamefont {Kj\"all}\ \emph {et~al.}(2014)\citenamefont
  {Kj\"all}, \citenamefont {Bardarson},\ and\ \citenamefont
  {Pollmann}}]{PhysRevLett.113.107204}%
  \BibitemOpen
  \bibfield  {author} {\bibinfo {author} {\bibfnamefont {J.~A.}\ \bibnamefont
  {Kj\"all}}, \bibinfo {author} {\bibfnamefont {J.~H.}\ \bibnamefont
  {Bardarson}}, \ and\ \bibinfo {author} {\bibfnamefont {F.}~\bibnamefont
  {Pollmann}},\ }\href {\doibase 10.1103/PhysRevLett.113.107204} {\bibfield
  {journal} {\bibinfo  {journal} {Phys. Rev. Lett.}\ }\textbf {\bibinfo
  {volume} {113}},\ \bibinfo {pages} {107204} (\bibinfo {year}
  {2014})}\BibitemShut {NoStop}%
\bibitem [{\citenamefont {Ng}\ and\ \citenamefont
  {Kolodrubetz}(2019)}]{mblapplication}%
  \BibitemOpen
  \bibfield  {author} {\bibinfo {author} {\bibfnamefont {N.}~\bibnamefont
  {Ng}}\ and\ \bibinfo {author} {\bibfnamefont {M.}~\bibnamefont
  {Kolodrubetz}},\ }\href {\doibase 10.1103/PhysRevLett.122.240402} {\bibfield
  {journal} {\bibinfo  {journal} {Phys. Rev. Lett.}\ }\textbf {\bibinfo
  {volume} {122}},\ \bibinfo {pages} {240402} (\bibinfo {year}
  {2019})}\BibitemShut {NoStop}%
\bibitem [{\citenamefont {Bohigas}\ \emph {et~al.}(1984)\citenamefont
  {Bohigas}, \citenamefont {Giannoni},\ and\ \citenamefont
  {Schmit}}]{PhysRevLett.52.1}%
  \BibitemOpen
  \bibfield  {author} {\bibinfo {author} {\bibfnamefont {O.}~\bibnamefont
  {Bohigas}}, \bibinfo {author} {\bibfnamefont {M.~J.}\ \bibnamefont
  {Giannoni}}, \ and\ \bibinfo {author} {\bibfnamefont {C.}~\bibnamefont
  {Schmit}},\ }\href {\doibase 10.1103/PhysRevLett.52.1} {\bibfield  {journal}
  {\bibinfo  {journal} {Phys. Rev. Lett.}\ }\textbf {\bibinfo {volume} {52}},\
  \bibinfo {pages} {1} (\bibinfo {year} {1984})}\BibitemShut {NoStop}%
\bibitem [{\citenamefont {Georgeot}\ and\ \citenamefont
  {Shepelyansky}(1998)}]{sgs}%
  \BibitemOpen
  \bibfield  {author} {\bibinfo {author} {\bibfnamefont {B.}~\bibnamefont
  {Georgeot}}\ and\ \bibinfo {author} {\bibfnamefont {D.~L.}\ \bibnamefont
  {Shepelyansky}},\ }\href {\doibase 10.1103/PhysRevLett.81.5129} {\bibfield
  {journal} {\bibinfo  {journal} {Phys. Rev. Lett.}\ }\textbf {\bibinfo
  {volume} {81}},\ \bibinfo {pages} {5129} (\bibinfo {year}
  {1998})}\BibitemShut {NoStop}%
\bibitem [{\citenamefont {Ericsson}\ and\ \citenamefont
  {Ruhe}(1980)}]{Thomas1980The}%
  \BibitemOpen
  \bibfield  {author} {\bibinfo {author} {\bibfnamefont {T.}~\bibnamefont
  {Ericsson}}\ and\ \bibinfo {author} {\bibfnamefont {A.}~\bibnamefont
  {Ruhe}},\ }\href@noop {} {\bibfield  {journal} {\bibinfo  {journal} {Math.
  Comput.}\ }\textbf {\bibinfo {volume} {35}},\ \bibinfo {pages} {1251}
  (\bibinfo {year} {1980})}\BibitemShut {NoStop}%
\bibitem [{\citenamefont {Wyatt}(1995)}]{PhysRevE.51.3643}%
  \BibitemOpen
  \bibfield  {author} {\bibinfo {author} {\bibfnamefont {R.~E.}\ \bibnamefont
  {Wyatt}},\ }\href {\doibase 10.1103/PhysRevE.51.3643} {\bibfield  {journal}
  {\bibinfo  {journal} {Phys. Rev. E}\ }\textbf {\bibinfo {volume} {51}},\
  \bibinfo {pages} {3643} (\bibinfo {year} {1995})}\BibitemShut {NoStop}%
\bibitem [{\citenamefont {Minehardt}\ \emph {et~al.}(1997)\citenamefont
  {Minehardt}, \citenamefont {Adcock},\ and\ \citenamefont
  {Wyatt}}]{PhysRevE.56.4837}%
  \BibitemOpen
  \bibfield  {author} {\bibinfo {author} {\bibfnamefont {T.~J.}\ \bibnamefont
  {Minehardt}}, \bibinfo {author} {\bibfnamefont {J.~D.}\ \bibnamefont
  {Adcock}}, \ and\ \bibinfo {author} {\bibfnamefont {R.~E.}\ \bibnamefont
  {Wyatt}},\ }\href {\doibase 10.1103/PhysRevE.56.4837} {\bibfield  {journal}
  {\bibinfo  {journal} {Phys. Rev. E}\ }\textbf {\bibinfo {volume} {56}},\
  \bibinfo {pages} {4837} (\bibinfo {year} {1997})}\BibitemShut {NoStop}%
\bibitem [{\citenamefont {Bai}\ \emph {et~al.}(2000)\citenamefont {Bai},
  \citenamefont {Demmel}, \citenamefont {Dongarra}, \citenamefont {Ruhe},\ and\
  \citenamefont {van~der Vorst~(Eds.)}}]{TSA}%
  \BibitemOpen
  \bibfield  {author} {\bibinfo {author} {\bibfnamefont {Z.}~\bibnamefont
  {Bai}}, \bibinfo {author} {\bibfnamefont {J.}~\bibnamefont {Demmel}},
  \bibinfo {author} {\bibfnamefont {J.}~\bibnamefont {Dongarra}}, \bibinfo
  {author} {\bibfnamefont {A.}~\bibnamefont {Ruhe}}, \ and\ \bibinfo {author}
  {\bibfnamefont {H.}~\bibnamefont {van~der Vorst~(Eds.)}},\ }\href@noop {}
  {\emph {\bibinfo {title} {Templates for the Solution of Algebraic Eigenvalue
  Problems: A Practical Guide.}}}\ (\bibinfo  {publisher} {SIAM,
  Philadelphia},\ \bibinfo {year} {2000})\BibitemShut {NoStop}%
\bibitem [{\citenamefont {Pietracaprina}\ \emph {et~al.}(2018)\citenamefont
  {Pietracaprina}, \citenamefont {Macé}, \citenamefont {Luitz},\ and\
  \citenamefont {Alet}}]{scipost}%
  \BibitemOpen
  \bibfield  {author} {\bibinfo {author} {\bibfnamefont {F.}~\bibnamefont
  {Pietracaprina}}, \bibinfo {author} {\bibfnamefont {N.}~\bibnamefont
  {Macé}}, \bibinfo {author} {\bibfnamefont {D.~J.}\ \bibnamefont {Luitz}}, \
  and\ \bibinfo {author} {\bibfnamefont {F.}~\bibnamefont {Alet}},\ }\href
  {\doibase 10.21468/SciPostPhys.5.5.045} {\bibfield  {journal} {\bibinfo
  {journal} {SciPost Phys.}\ }\textbf {\bibinfo {volume} {5}},\ \bibinfo
  {pages} {45} (\bibinfo {year} {2018})}\BibitemShut {NoStop}%
\bibitem [{\citenamefont {Davidson}(1989)}]{Davidson1989Super}%
  \BibitemOpen
  \bibfield  {author} {\bibinfo {author} {\bibfnamefont {E.~R.}\ \bibnamefont
  {Davidson}},\ }\href@noop {} {\bibfield  {journal} {\bibinfo  {journal}
  {Comput. Phys. Commun.}\ }\textbf {\bibinfo {volume} {53}},\ \bibinfo {pages}
  {49} (\bibinfo {year} {1989})}\BibitemShut {NoStop}%
\bibitem [{\citenamefont {Saad}(2011)}]{YSaad}%
  \BibitemOpen
  \bibfield  {author} {\bibinfo {author} {\bibfnamefont {Y.}~\bibnamefont
  {Saad}},\ }\href@noop {} {\emph {\bibinfo {title} {Numerical Methods for
  Large Eigenvalue Problems}}}\ (\bibinfo  {publisher} {SIAM, Philadelphia},\
  \bibinfo {year} {2011})\BibitemShut {NoStop}%
\bibitem [{\citenamefont {Dorando}\ \emph {et~al.}(2007)\citenamefont
  {Dorando}, \citenamefont {Hachmann},\ and\ \citenamefont
  {Chan}}]{Dorando2007Targeted}%
  \BibitemOpen
  \bibfield  {author} {\bibinfo {author} {\bibfnamefont {J.~J.}\ \bibnamefont
  {Dorando}}, \bibinfo {author} {\bibfnamefont {J.}~\bibnamefont {Hachmann}}, \
  and\ \bibinfo {author} {\bibfnamefont {G.~K.-L.}\ \bibnamefont {Chan}},\
  }\href@noop {} {\bibfield  {journal} {\bibinfo  {journal} {J. Chem. Phys.}\
  }\textbf {\bibinfo {volume} {127}},\ \bibinfo {pages} {401} (\bibinfo {year}
  {2007})}\BibitemShut {NoStop}%
\bibitem [{\citenamefont {Jordan}\ \emph {et~al.}(2012)\citenamefont {Jordan},
  \citenamefont {Marsman}, \citenamefont {Kim},\ and\ \citenamefont
  {Kresse}}]{Jordan2012Fast}%
  \BibitemOpen
  \bibfield  {author} {\bibinfo {author} {\bibfnamefont {G.}~\bibnamefont
  {Jordan}}, \bibinfo {author} {\bibfnamefont {M.}~\bibnamefont {Marsman}},
  \bibinfo {author} {\bibfnamefont {Y.-S.}\ \bibnamefont {Kim}}, \ and\
  \bibinfo {author} {\bibfnamefont {G.}~\bibnamefont {Kresse}},\ }\href@noop {}
  {\bibfield  {journal} {\bibinfo  {journal} {J. Comput. Phys.}\ }\textbf
  {\bibinfo {volume} {231}},\ \bibinfo {pages} {4836} (\bibinfo {year}
  {2012})}\BibitemShut {NoStop}%
\bibitem [{\citenamefont {Neuhauser}(1990)}]{Neuhauser1990Bound}%
  \BibitemOpen
  \bibfield  {author} {\bibinfo {author} {\bibfnamefont {D.}~\bibnamefont
  {Neuhauser}},\ }\href@noop {} {\bibfield  {journal} {\bibinfo  {journal} {J.
  Chem. Phys.}\ }\textbf {\bibinfo {volume} {93}},\ \bibinfo {pages} {2611}
  (\bibinfo {year} {1990})}\BibitemShut {NoStop}%
\bibitem [{\citenamefont {Neuhauser}(1991)}]{Neuhauser1991Time}%
  \BibitemOpen
  \bibfield  {author} {\bibinfo {author} {\bibfnamefont {D.}~\bibnamefont
  {Neuhauser}},\ }\href@noop {} {\bibfield  {journal} {\bibinfo  {journal} {J.
  Chem. Phys.}\ }\textbf {\bibinfo {volume} {95}},\ \bibinfo {pages} {4927}
  (\bibinfo {year} {1991})}\BibitemShut {NoStop}%
\bibitem [{\citenamefont {Santra}\ \emph {et~al.}(2000)\citenamefont {Santra},
  \citenamefont {Breidbach}, \citenamefont {Zobeley},\ and\ \citenamefont
  {Cederbaum}}]{Santra2000Parallel}%
  \BibitemOpen
  \bibfield  {author} {\bibinfo {author} {\bibfnamefont {R.}~\bibnamefont
  {Santra}}, \bibinfo {author} {\bibfnamefont {J.}~\bibnamefont {Breidbach}},
  \bibinfo {author} {\bibfnamefont {J.}~\bibnamefont {Zobeley}}, \ and\
  \bibinfo {author} {\bibfnamefont {L.~S.}\ \bibnamefont {Cederbaum}},\
  }\href@noop {} {\bibfield  {journal} {\bibinfo  {journal} {J. Chem. Phys.}\
  }\textbf {\bibinfo {volume} {112}},\ \bibinfo {pages} {9243} (\bibinfo {year}
  {2000})}\BibitemShut {NoStop}%
\bibitem [{\citenamefont {Vijay}\ and\ \citenamefont
  {Wyatt}(2000)}]{PhysRevE.62.4351}%
  \BibitemOpen
  \bibfield  {author} {\bibinfo {author} {\bibfnamefont {A.}~\bibnamefont
  {Vijay}}\ and\ \bibinfo {author} {\bibfnamefont {R.~E.}\ \bibnamefont
  {Wyatt}},\ }\href {\doibase 10.1103/PhysRevE.62.4351} {\bibfield  {journal}
  {\bibinfo  {journal} {Phys. Rev. E}\ }\textbf {\bibinfo {volume} {62}},\
  \bibinfo {pages} {4351} (\bibinfo {year} {2000})}\BibitemShut {NoStop}%
\bibitem [{\citenamefont {Pieper}\ \emph {et~al.}(2016)\citenamefont {Pieper},
  \citenamefont {Kreutzer}, \citenamefont {Alvermann}, \citenamefont {Galgon},
  \citenamefont {Fehske}, \citenamefont {Hager}, \citenamefont {Lang},\ and\
  \citenamefont {Wellein}}]{Pieper2016High}%
  \BibitemOpen
  \bibfield  {author} {\bibinfo {author} {\bibfnamefont {A.}~\bibnamefont
  {Pieper}}, \bibinfo {author} {\bibfnamefont {M.}~\bibnamefont {Kreutzer}},
  \bibinfo {author} {\bibfnamefont {A.}~\bibnamefont {Alvermann}}, \bibinfo
  {author} {\bibfnamefont {M.}~\bibnamefont {Galgon}}, \bibinfo {author}
  {\bibfnamefont {H.}~\bibnamefont {Fehske}}, \bibinfo {author} {\bibfnamefont
  {G.}~\bibnamefont {Hager}}, \bibinfo {author} {\bibfnamefont
  {B.}~\bibnamefont {Lang}}, \ and\ \bibinfo {author} {\bibfnamefont
  {G.}~\bibnamefont {Wellein}},\ }\href@noop {} {\bibfield  {journal} {\bibinfo
   {journal} {J. Comput. Phys.}\ }\textbf {\bibinfo {volume} {325}},\ \bibinfo
  {pages} {226} (\bibinfo {year} {2016})}\BibitemShut {NoStop}%
\bibitem [{\citenamefont {Fang}\ and\ \citenamefont {Saad}(2012)}]{AFLP}%
  \BibitemOpen
  \bibfield  {author} {\bibinfo {author} {\bibfnamefont {H.}~\bibnamefont
  {Fang}}\ and\ \bibinfo {author} {\bibfnamefont {Y.}~\bibnamefont {Saad}},\
  }\href@noop {} {\bibfield  {journal} {\bibinfo  {journal} {SIAM J. Sci.
  Comput.}\ }\textbf {\bibinfo {volume} {34}},\ \bibinfo {pages} {A2220}
  (\bibinfo {year} {2012})}\BibitemShut {NoStop}%
\bibitem [{\citenamefont {Li}\ \emph {et~al.}(2016)\citenamefont {Li},
  \citenamefont {Xi}, \citenamefont {Vecharynski}, \citenamefont {Yang},\ and\
  \citenamefont {Saad}}]{TRA}%
  \BibitemOpen
  \bibfield  {author} {\bibinfo {author} {\bibfnamefont {R.}~\bibnamefont
  {Li}}, \bibinfo {author} {\bibfnamefont {Y.}~\bibnamefont {Xi}}, \bibinfo
  {author} {\bibfnamefont {E.}~\bibnamefont {Vecharynski}}, \bibinfo {author}
  {\bibfnamefont {C.}~\bibnamefont {Yang}}, \ and\ \bibinfo {author}
  {\bibfnamefont {Y.}~\bibnamefont {Saad}},\ }\href@noop {} {\bibfield
  {journal} {\bibinfo  {journal} {SIAM J. Sci. Comput.}\ }\textbf {\bibinfo
  {volume} {38}},\ \bibinfo {pages} {A2512} (\bibinfo {year}
  {2016})}\BibitemShut {NoStop}%
\bibitem [{\citenamefont {Zhou}(2010)}]{cd2}%
  \BibitemOpen
  \bibfield  {author} {\bibinfo {author} {\bibfnamefont {Y.}~\bibnamefont
  {Zhou}},\ }\href {\doibase 10.1016/j.jcp.2010.08.032} {\bibfield  {journal}
  {\bibinfo  {journal} {J. Comput. Phys.}\ }\textbf {\bibinfo {volume} {229}},\
  \bibinfo {pages} {9188} (\bibinfo {year} {2010})}\BibitemShut {NoStop}%
\bibitem [{\citenamefont {Zhou}\ and\ \citenamefont {Saad}(2007)}]{cd1}%
  \BibitemOpen
  \bibfield  {author} {\bibinfo {author} {\bibfnamefont {Y.}~\bibnamefont
  {Zhou}}\ and\ \bibinfo {author} {\bibfnamefont {Y.}~\bibnamefont {Saad}},\
  }\href@noop {} {\bibfield  {journal} {\bibinfo  {journal} {SIAM J. Matrix
  Anal. Appl.}\ }\textbf {\bibinfo {volume} {29}},\ \bibinfo {pages} {954}
  (\bibinfo {year} {2007})}\BibitemShut {NoStop}%
\bibitem [{\citenamefont {Mason}\ and\ \citenamefont
  {Handscomb}(2003)}]{Handscomb2003Chebyshev}%
  \BibitemOpen
  \bibfield  {author} {\bibinfo {author} {\bibfnamefont {J.~C.}\ \bibnamefont
  {Mason}}\ and\ \bibinfo {author} {\bibfnamefont {D.~C.}\ \bibnamefont
  {Handscomb}},\ }\href@noop {} {\emph {\bibinfo {title} {Chebyshev
  polynomials}}}\ (\bibinfo  {publisher} {CRC Press LLC, Boca Raton},\ \bibinfo
  {year} {2003})\BibitemShut {NoStop}%
\bibitem [{\citenamefont {Boyd}(2000)}]{ChebandFourier}%
  \BibitemOpen
  \bibfield  {author} {\bibinfo {author} {\bibfnamefont {J.~P.}\ \bibnamefont
  {Boyd}},\ }\href@noop {} {\emph {\bibinfo {title} {Chebyshev and Fourier
  Spectral Methods}}}\ (\bibinfo  {publisher} {Dover Publications, New York},\
  \bibinfo {year} {2000})\BibitemShut {NoStop}%
\bibitem [{\citenamefont {Georgeot}\ and\ \citenamefont
  {Shepelyansky}(2000)}]{PhysRevE.62.3504}%
  \BibitemOpen
  \bibfield  {author} {\bibinfo {author} {\bibfnamefont {B.}~\bibnamefont
  {Georgeot}}\ and\ \bibinfo {author} {\bibfnamefont {D.~L.}\ \bibnamefont
  {Shepelyansky}},\ }\href {\doibase 10.1103/PhysRevE.62.3504} {\bibfield
  {journal} {\bibinfo  {journal} {Phys. Rev. E}\ }\textbf {\bibinfo {volume}
  {62}},\ \bibinfo {pages} {3504} (\bibinfo {year} {2000})}\BibitemShut
  {NoStop}%
\bibitem [{\citenamefont {Daniel}\ \emph {et~al.}(1976)\citenamefont {Daniel},
  \citenamefont {Gragg}, \citenamefont {Kaufman},\ and\ \citenamefont
  {W.}}]{10.2307/2005398}%
  \BibitemOpen
  \bibfield  {author} {\bibinfo {author} {\bibfnamefont {J.~W.}\ \bibnamefont
  {Daniel}}, \bibinfo {author} {\bibfnamefont {W.~B.}\ \bibnamefont {Gragg}},
  \bibinfo {author} {\bibfnamefont {L.}~\bibnamefont {Kaufman}}, \ and\
  \bibinfo {author} {\bibfnamefont {G.}~\bibnamefont {W.}},\ }\href
  {http://www.jstor.org/stable/2005398} {\bibfield  {journal} {\bibinfo
  {journal} {Math. Comput.}\ }\textbf {\bibinfo {volume} {30}},\ \bibinfo
  {pages} {772} (\bibinfo {year} {1976})}\BibitemShut {NoStop}%
\bibitem [{\citenamefont {Zhou}\ and\ \citenamefont {Li}(2011)}]{ZHOU2011480}%
  \BibitemOpen
  \bibfield  {author} {\bibinfo {author} {\bibfnamefont {Y.}~\bibnamefont
  {Zhou}}\ and\ \bibinfo {author} {\bibfnamefont {R.-C.}\ \bibnamefont {Li}},\
  }\href {\doibase https://doi.org/10.1016/j.laa.2010.06.034} {\bibfield
  {journal} {\bibinfo  {journal} {Linear Algebra Appl.}\ }\textbf {\bibinfo
  {volume} {435}},\ \bibinfo {pages} {480 } (\bibinfo {year}
  {2011})}\BibitemShut {NoStop}%
\bibitem [{\citenamefont {Lehoucq}\ \emph {et~al.}(1998)\citenamefont
  {Lehoucq}, \citenamefont {Sorensen},\ and\ \citenamefont {Yang}}]{Arpack}%
  \BibitemOpen
  \bibfield  {author} {\bibinfo {author} {\bibfnamefont {R.~B.}\ \bibnamefont
  {Lehoucq}}, \bibinfo {author} {\bibfnamefont {D.~C.}\ \bibnamefont
  {Sorensen}}, \ and\ \bibinfo {author} {\bibfnamefont {C.}~\bibnamefont
  {Yang}},\ }\href@noop {} {\emph {\bibinfo {title} {ARPACK User's Guide:
  Solution of Large Scale Eigenvalue Problems with Implicitly Restarted Arnoldi
  Methods}}}\ (\bibinfo  {publisher} {SIAM, Philadelphia},\ \bibinfo {year}
  {1998})\BibitemShut {NoStop}%
\bibitem [{\citenamefont {Hams}\ and\ \citenamefont
  {De~Raedt}(2000)}]{dosestimation}%
  \BibitemOpen
  \bibfield  {author} {\bibinfo {author} {\bibfnamefont {A.}~\bibnamefont
  {Hams}}\ and\ \bibinfo {author} {\bibfnamefont {H.}~\bibnamefont
  {De~Raedt}},\ }\href {\doibase 10.1103/PhysRevE.62.4365} {\bibfield
  {journal} {\bibinfo  {journal} {Phys. Rev. E}\ }\textbf {\bibinfo {volume}
  {62}},\ \bibinfo {pages} {4365} (\bibinfo {year} {2000})}\BibitemShut
  {NoStop}%
\bibitem [{\citenamefont {Fisher}(1995)}]{PhysRevB.51.6411}%
  \BibitemOpen
  \bibfield  {author} {\bibinfo {author} {\bibfnamefont {D.~S.}\ \bibnamefont
  {Fisher}},\ }\href {\doibase 10.1103/PhysRevB.51.6411} {\bibfield  {journal}
  {\bibinfo  {journal} {Phys. Rev. B}\ }\textbf {\bibinfo {volume} {51}},\
  \bibinfo {pages} {6411} (\bibinfo {year} {1995})}\BibitemShut {NoStop}%
\bibitem [{\citenamefont {Schollw\"ock}(2005)}]{RevModPhys.77.259}%
  \BibitemOpen
  \bibfield  {author} {\bibinfo {author} {\bibfnamefont {U.}~\bibnamefont
  {Schollw\"ock}},\ }\href {\doibase 10.1103/RevModPhys.77.259} {\bibfield
  {journal} {\bibinfo  {journal} {Rev. Mod. Phys.}\ }\textbf {\bibinfo {volume}
  {77}},\ \bibinfo {pages} {259} (\bibinfo {year} {2005})}\BibitemShut
  {NoStop}%
\bibitem [{\citenamefont {Anderson}\ \emph {et~al.}(1999)\citenamefont
  {Anderson}, \citenamefont {Bai}, \citenamefont {Bischof}, \citenamefont
  {Blackford}, \citenamefont {Demmel}, \citenamefont {Dongarra}, \citenamefont
  {Croz}, \citenamefont {Hammarling}, \citenamefont {Greenbaum},\ and\
  \citenamefont {Mckenney}}]{Anderson1999LAPACK}%
  \BibitemOpen
  \bibfield  {author} {\bibinfo {author} {\bibfnamefont {E.}~\bibnamefont
  {Anderson}}, \bibinfo {author} {\bibfnamefont {Z.}~\bibnamefont {Bai}},
  \bibinfo {author} {\bibfnamefont {C.}~\bibnamefont {Bischof}}, \bibinfo
  {author} {\bibfnamefont {L.~S.}\ \bibnamefont {Blackford}}, \bibinfo {author}
  {\bibfnamefont {J.}~\bibnamefont {Demmel}}, \bibinfo {author} {\bibfnamefont
  {J.~J.}\ \bibnamefont {Dongarra}}, \bibinfo {author} {\bibfnamefont {J.~D.}\
  \bibnamefont {Croz}}, \bibinfo {author} {\bibfnamefont {S.}~\bibnamefont
  {Hammarling}}, \bibinfo {author} {\bibfnamefont {A.}~\bibnamefont
  {Greenbaum}}, \ and\ \bibinfo {author} {\bibfnamefont {A.}~\bibnamefont
  {Mckenney}},\ }\href@noop {} {\emph {\bibinfo {title} {LAPACK Users' guide
  (third ed.)}}}\ (\bibinfo  {publisher} {SIAM, Philadelphia},\ \bibinfo {year}
  {1999})\BibitemShut {NoStop}%
\bibitem [{\citenamefont {Sorensen}(1992)}]{implicit}%
  \BibitemOpen
  \bibfield  {author} {\bibinfo {author} {\bibfnamefont {D.~C.}\ \bibnamefont
  {Sorensen}},\ }\href@noop {} {\bibfield  {journal} {\bibinfo  {journal} {SIAM
  J. Matrix Anal. Appl.}\ }\textbf {\bibinfo {volume} {13}},\ \bibinfo {pages}
  {357} (\bibinfo {year} {1992})}\BibitemShut {NoStop}%
\bibitem [{\citenamefont {Stewart}(2001)}]{Stewart}%
  \BibitemOpen
  \bibfield  {author} {\bibinfo {author} {\bibfnamefont {G.~W.}\ \bibnamefont
  {Stewart}},\ }\href@noop {} {\bibfield  {journal} {\bibinfo  {journal} {SIAM
  J. Matrix Anal. Appl.}\ }\textbf {\bibinfo {volume} {23}},\ \bibinfo {pages}
  {601} (\bibinfo {year} {2001})}\BibitemShut {NoStop}%
\bibitem [{\citenamefont {Luitz}\ \emph {et~al.}(2015)\citenamefont {Luitz},
  \citenamefont {Laflorencie},\ and\ \citenamefont
  {Alet}}]{PhysRevB.91.081103}%
  \BibitemOpen
  \bibfield  {author} {\bibinfo {author} {\bibfnamefont {D.~J.}\ \bibnamefont
  {Luitz}}, \bibinfo {author} {\bibfnamefont {N.}~\bibnamefont {Laflorencie}},
  \ and\ \bibinfo {author} {\bibfnamefont {F.}~\bibnamefont {Alet}},\ }\href
  {\doibase 10.1103/PhysRevB.91.081103} {\bibfield  {journal} {\bibinfo
  {journal} {Phys. Rev. B}\ }\textbf {\bibinfo {volume} {91}},\ \bibinfo
  {pages} {081103} (\bibinfo {year} {2015})}\BibitemShut {NoStop}%
\bibitem [{\citenamefont {Sierant}\ and\ \citenamefont
  {Zakrzewski}(2020)}]{PhysRevB.101.104201}%
  \BibitemOpen
  \bibfield  {author} {\bibinfo {author} {\bibfnamefont {P.}~\bibnamefont
  {Sierant}}\ and\ \bibinfo {author} {\bibfnamefont {J.}~\bibnamefont
  {Zakrzewski}},\ }\href {\doibase 10.1103/PhysRevB.101.104201} {\bibfield
  {journal} {\bibinfo  {journal} {Phys. Rev. B}\ }\textbf {\bibinfo {volume}
  {101}},\ \bibinfo {pages} {104201} (\bibinfo {year} {2020})}\BibitemShut
  {NoStop}%
\bibitem [{\citenamefont {Hopjan}\ and\ \citenamefont
  {Heidrich-Meisner}(2020)}]{PhysRevA.101.063617}%
  \BibitemOpen
  \bibfield  {author} {\bibinfo {author} {\bibfnamefont {M.}~\bibnamefont
  {Hopjan}}\ and\ \bibinfo {author} {\bibfnamefont {F.}~\bibnamefont
  {Heidrich-Meisner}},\ }\href {\doibase 10.1103/PhysRevA.101.063617}
  {\bibfield  {journal} {\bibinfo  {journal} {Phys. Rev. A}\ }\textbf {\bibinfo
  {volume} {101}},\ \bibinfo {pages} {063617} (\bibinfo {year}
  {2020})}\BibitemShut {NoStop}%
\bibitem [{\citenamefont {Montangero}(2018)}]{ITNMN}%
  \BibitemOpen
  \bibfield  {author} {\bibinfo {author} {\bibfnamefont {S.}~\bibnamefont
  {Montangero}},\ }\href@noop {} {\emph {\bibinfo {title} {Introduction to
  Tensor Network Methods: Numerical simulations of low-dimensional many-body
  quantum systems}}},\ \bibinfo {edition} {1st}\ ed.\ (\bibinfo  {publisher}
  {Springer International Publishing},\ \bibinfo {year} {2018})\BibitemShut
  {NoStop}%
\end{thebibliography}%
\providecommand{\noopsort}[1]{}\providecommand{\singleletter}[1]{#1}%

\end{document}